\documentclass[aip,graphicx]{revtex4-1}
\usepackage{graphicx}
\draft 

\newcommand{\bea}{\begin{eqnarray}}
\newcommand{\eea}{\end{eqnarray}}
\newcommand{\be}{\begin{equation}}
\newcommand{\ee}{\end{equation}}
\newcommand{\dd}{\mbox{d}}
\newcommand{\ii}{{\rm i}}

\begin{document}

\title{The study of the dynamics of the order parameter of coupled oscillators in the Ott-Antonsen scheme
for generic frequency distributions}

\author{Alessandro Campa}
\email{alessandro.campa@iss.it}
\affiliation{National Center for Radiation Protection and
Computational Physics, Istituto Superiore di Sanit\`{a},
Viale Regina Elena 299, 00161 Rome, Italy}

\date{\today}

\begin{abstract}
The Ott-Antonsen ansatz shows that, for certain classes of distribution of the natural frequencies in systems
of $N$ globally coupled Kuramoto oscillators, the dynamics of the order parameter, in the limit $N\to \infty$, evolves,
under suitable initial conditions, in a manifold of low dimension. This is not possible when the frequency
distribution, continued in the complex plane, has an essential singularity at infinite; this is the case for
example, of a Gaussian distribution. In this work we propose a simple approximation scheme that allows to extend
also to this case the representation of the dynamics of the order parameter in a low dimensional manifold. Using
as a working example the Gaussian frequency distribution, we compare the dynamical evolution of the order parameter
of the system of oscillators, obtained by the numerical integration of the $N$ equations of motion, with the analogous
dynamics in the low dimensional manifold obtained with the application of the approximation scheme. The results confirm the
validity of the approximation. The method could be employed for general frequency distributions, allowing the
determination of the corresponding phase diagram of the oscillator system.
\end{abstract}


\maketitle

\begin{quotation}
Large systems of globally coupled oscillators with distributed natural frequencies can show a variety of
different behaviors, depending on the strength of their interaction and on the properties of the
distribution function of the frequencies. They can settle asymptotically to incoherent states, to
partially synchronized states, and also to periodic states. The availability of a rapid way to
study the dynamics of the system without the necessity to numerically integrate the equations of motion
(that obviously can be time demanding when the number of oscillators is very large)
can allow to determine easily when these different situations occur, varying the values of the parameters
of the frequency distribution and of the coupling constant between the oscillators; one could then obtain the phase
diagram of the oscillator system. Such a rapid way is possible if the dynamics of the order parameter evolves in a low dimensional
manifold and can be described by few nonlinear ordinary differential equations. This occurs applying the
Ott-Antonsen ansatz \cite{ottant} when the frequency distribution can be analytically continued to complex
frequencies to a function with only few poles in the lower complex plane. In this work we introduce an
approximation procedure to obtain a systems of few differential equations also for frequencies distributions,
like the Gaussian, with essential singularities in the complex plane. The analysis of the order parameter dynamics,
evaluated by numerically solving the equations in the low dimensional manifold, shows that the approximation
can be useful to compute the properties of oscillator systems with generic distributions of natural frequencies.
\end{quotation}

\section{Introduction}
\label{secintro}
The realization of the great importance of collective synchronization in systems of very different nature,
belonging to the realm of Physics and to that of Biology \cite{pikovskybook}, has caused a surge of interest and of activity
to understand more and more deeply the mechanisms at the basis of the phenomenon. In the many different examples
that it is possible to cite, like voltage oscillations in arrays of Josephson junctions \cite{wiesenfeld1998}, flashing in groups
of fireflies \cite{buck1988}, firing of cardiac pacemaker cells \cite{winfree1980}, phases of power units in electrical power distribution
networks \cite{filatrella2008}, metabolism in yeast cell suspensions \cite{bier2000}, and more others \cite{pikovskybook,strogatzbook},
we can find the common property of the emergence of a cooperative effect in which basic units, characterized by different
natural frequencies, spontaneously synchronize, to oscillate at a common frequency.

A simplified description of these systems can be constructed using systems of phase-only oscillators that interact globally.
Although such description neglects many details of the real systems, nevertheless it is able to grasp the essential
details of the synchronization mechanism. The celebrated model introduced by Kuramoto \cite{kuraorig} can be considered the
simplest one, as far as the interaction among the oscillators is concerned. In fact, given that the interaction between two
oscillators should depend on the difference of their phases, and that a function of this difference could be naturally
expressed as a Fourier series containing only the sines (in order to have an interaction in which the torque
on the $i$-th oscillator exterted by the $j$-th oscillator is the opposite of that exerted on the $j$-th oscillator by the $i$-th
oscillator), the Kuramoto model assumes that only the first Fourier component is present. Then, the dynamics of $N$ fully
coupled oscillators is described, in the Kuramoto model, by the following system of equations \cite{kuraorig}:
\be
\frac{\dd \theta_i}{\dd t} = \omega_i -\frac{K}{N}\sum_{j=1}^N \sin \left( \theta_i - \theta_j \right)
\, , \,\,\,\,\,\,\,\,\,\,\,\,\,\,\,\,\,\, i=1,\dots,N, \,
\label{eq_kura_mod}
\ee
where $\theta_i \in [0,2\pi)$ is the phase of the $i$-th oscillator and $\omega_i$ is its natural frequency;
the positive parameter $K$ is the coupling constant among the oscillators, normalized by the number $N$ of oscillators. The scaling
of the coupling with $N$ means that each oscillator is weakly coupled to all other oscillators \cite{kuraorig}, and it
allows to have a well defined behavior \cite{physrep} in the limit $N \to \infty$. The values of the natural
frequencies $\omega_i$ are extracted by a given distribution function $g(\omega)$. A very important quantity
is the complex order parameter $r(t)$, defined by
\be
\label{deforder_disc}
r(t) = \frac{1}{N} \sum_{i=1}^N e^{\ii \theta_i(t)} \, .
\ee
By its definition, the modulus of the order parameter is between $0$ and $1$. The qualitative properties of the dynamics
can be associated to the long time behavior of this order parameter. This problem
is typically considered in the limit $N \to \infty$.
More precisely, the dynamics of the system is considered in the limit $N\to \infty$, in which, as recalled later and detailed
in the next section, it is described by the evolution of a distribution function. Then one studies, in this framework, the long time
limit (formally $t\to \infty$) of this dynamics.
Depending on the value of the coupling constant $K$ the asymptotic properties
of $r(t)$ are different. Since the interaction between the oscillators is attractive, it is reasonable to expect that for
sufficiently large $K$ the oscillators tend to partially synchronize, giving rise to a finite modulus $R(t) \equiv |r(t)|$ of the order
parameter, while for small $K$ the interactions is not sufficient to overcome the dispersion caused by the different natural
frequencies, and the oscillator phases tend to spread uniformly between $0$ and $2\pi$, resulting in a vanishing $R(t)$, namely
the oscillators display an incoherent state. The details
depend on the characteristics of the frequency distribution $g(\omega)$. One can represent the long-time properties
in a phase diagram, i.e., a diagram having for coordinates quantities depending on the values of the parameters of the frequency
distribution $g(\omega)$ and of $K$; the diagram will be divided in regions characterized by the different asymptotic behaviors of the order
parameter \cite{footnote1}. The structure of
the phase diagram will be more or less elaborate depending on the form of $g(\omega)$, possibly with regions where $R(t)$ does not
even settle to a stationary value, but the system reaches a state in which $R(t)$ oscillates. The simplest case arises with a unimodal
symmetric distribution, i.e., when $g(\omega)$ has a single maximum in $\omega=0$ and $g(-\omega) = g(\omega)$. Then for $K$ less than
a critical value $K_c$, which is equal to $2/(\pi g(0))$, the system settles asymptotically to an incoherent state,
with $R=0$; at $K=K_c$ there is a continuous transition to a partially synchronized state with $R$ settling to a constant value,
increasing with $K$ and tending to $1$ for $K \to \infty$. This picture would hold in the more general case of
a unimodal frequency distribution centered at a frequency $\omega_0 \ne 0$ and symmetric with respect to $\omega_0$. This could be seen
by simply going to a comoving frame rotating with freqeuncy $\omega_0$; in this frame, the phase diagram would be as the one just described.

Then the state of synchrony, or absence of synchrony, is described by the asymptotic states of the system of oscillators.
For generic frequency distributions $g(\omega)$ the determination of the phase diagram is not as simple as for unimodal symmetric
distributions, when the features mentioned above can be readily proved \cite{strogatz2000,briefbook}. The simulation of the
dnamics, i.e., the numerical integration of the equations of motion (\ref{eq_kura_mod}) for very large $N$, would provide the
necessary information. However, this would be highly impractical, since the exploration of the different regions of the phase
diagram would require a very large computing time. Therefore, a procedure that approximates the full dynamics of the order
parameter with one evolving
in a manifold of low dimension, and thus determined by few differential equations, would greatly simplify the task.

In the limit $N\to \infty$ the dynamics is described by the
one-particle distribution function $\rho(\theta,\omega,t)$, with $\rho(\theta,\omega,t) \dd \theta$ proportional, for each $\omega$,
to the fraction of oscillators, among those with proper frequency $\omega$, that at time $t$ have phases between
$\theta$ and $\theta + \dd \theta$. Consequently, the quantity $g(\omega)\rho(\theta,\omega,t) \dd \theta \dd \omega$ is
proportional to the fraction of oscillators, among the whole population, that have frequencies between $\omega$ and $\omega + \dd \omega$
and that at time $t$ have phases between $\theta$ and $\theta + \dd \theta$. 

The partial differential equation governing the evolution of $\rho(\theta,\omega,t)$ will be given in the next section.
This dynamics, or more precisely the dynamics of the complex order parameter $r(t)$, is the object of interest in this work. 
Going in Fourier space, the distribution $\rho(\theta,\omega,t)$ is replaced by its Fourier components $f_n(\omega,t)$, where,
$\rho$ being real, it is possible to consider only $n\ge 0$; the dynamics is then governed by a system
of coupled partial differential equations for $f_n(\omega,t)$. Thus, from the double continuous infinity, $\theta$ and $\omega$,
characterizing the dynamics of $\rho(\theta,\omega,t)$, one passes to a countable infinity, $n$, plus a continuous infinity,
$\omega$, for the description of the dynamics of the Fourier components.

The very important step forward made in Ref. \onlinecite{ottant} was the introduction of the assumption
$f_n(\omega,t)=f_1^n(\omega,t)$ for the Fourier components, the Ott-Antonsen (OA) ansatz, allowing, as shown in the
next section, to have a closed equation for $f_1(\omega,t)$. In this work we do not touch the problem
of the justification of the ansatz, and we will be concerned only with the dynamics of $f_1(\omega,t)$ as determined
{\it within} the OA ansatz. Comments and remarks will be given in the concluding Section.
In order to have as much as possible a self-contained presentation, a brief description of the OA procedure will be
given. In spite of the simplification introduced by the ansatz, the continuous infinity
related to the dependence of $f_1(\omega,t)$ on $\omega$ is still present. The great practical value of the
OA ansatz is revealed when the distribution function $g(\omega)$ of the oscillators' proper
frequencies can be analytically continued in the complex $\omega$ plane and the singularities of this
continuation are constituted by only few poles in the lower half plane, while in addition $g(\omega)$ vanishes
for ${\rm Im} (\omega) \to -\infty$. Then, with few additional assumptions recalled in the next
section, the dynamics of the order parameter is described by a small number of ordinary differential equations, more precisely
by a complex equation or two real equations for each pole of the continuation. The order parameter of the system
of oscillators is simply related to these few dynamical variables. Thus, the description of the order parameter dynamics, that
within the OA ansatz is provided, for a general $g(\omega)$, by the time evolution of the Fourier
component $f_1(\omega,t)$, in this case is given by the time evolution of few variables: the dependence on the
continuous variable $\omega$ is completely avoided, making the study of the dynamics much more feasible.
These few variables are the coordinates of the low dimensional manifold alluded to above.
The reduction of the dynamics obtained by the OA ansatz, and in particular the reduction
to a low dimensional manifold, implies that the dynamics is not reproduced exactly. Nevertheless, {\it if one is interested in
the long-time behavior of the order parameter}, this reduction is able to get this behavior. In the discussion we will
comment on this point and on the relation between the full dynamics and the reduced dynamics.

This reduction cannot be obtained when the analytical continuation of $g(\omega)$ does not result in a complex
function with few poles in the lower half plane. In particular, for perhaps the most representative example
one can think of, the one in which $g(\omega)$ is proportional to a Gaussian, i.e.,
$g(\omega) \propto e^{-d \omega^2}$ (with $d$ a positive constant), the analytical continuation, while having no
singularity at any finite value of $\omega$, nevertheless has an essential singularity at infinite, diverging exponentially when
$|\omega| \to \infty$ with $|{\rm Im}(\omega)| > |{\rm Re}(\omega)|$.

On the other hand, on physical grounds one would not expect substantial differences in the dynamical behavior of
the system for frequencies distributions $g(\omega)$ that, numerically, are not very different.
With this, we come to the main focus of this work. Let us consider, for
example, a Lorentzian $g(\omega)$ and a Gaussian $g(\omega)$. It is true that the decay for large $\omega$
is qualitatively very different, like $\omega^{-2}$ in the first case and like $e^{-d \omega^2}$ in the second case,
but one could build two such functions in such a way that the fraction of oscillators with proper frequencies
between $\omega$ and $\omega + \dd \omega$ is numerically similar, at least for the great majority of the oscillators;
it will not be so for the outliers with large $|\omega|$, but we do not expect this to influence much the qualitative
features of the dynamics. As a fact that can possibly confirm this expectation, we could consider the general properties
of the stable stationary solutions of the partial differential equation describing the evolution of $\rho(\theta,\omega,t)$
in the case of a symmetric and unimodal distributions $g(\omega)$, solution associated to the asymtpotic state of the system
in which the modulus $R$ of the order parameter settles to a constant value. As anticipated above, we know that, independently
from the particular expression of any such distribution, the modulus $R(K)$ in the asymptotic
state as a function of the coupling constant $K$, has always the same behavior, starting
from zero at a threshold value $K_c$ and then monotonously increasing with $K$, tending to $1$ for $K \to \infty$.
The precise form of $R(K)$ depends on $g(\omega)$, but the qualitative behavior is the same.
Therefore, it would be useful to find a way to circumvent the mathematical impossibility to reduce the order parameter dynamics to that
of few dynamical variables in the cases in which the distribution $g(\omega)$ does not enjoy the properties mentioned above
concerning its analytical continuation in the complex plane. It is our purpose in this work to show that this can be
achieved at the price of introducing some approximations in the treatment of the distribution $g(\omega)$. Although we will present
our approximation scheme for the example of a Gaussian frequency distribution, it will be clear that it can be equally
applied to any distribution that does not have only some poles in the lower half of the complex $\omega$ plane.

The paper is structured as follows. Section \ref{secmethod} is divided in two parts. In Section \ref{subseckura} we provide
a brief recalling of the OA method, in order to have, as remarked above, a self-contained presentation. Furthermore,
in this first part of Section \ref{secmethod} we will dedicate some space to an analysis of the role played by the initial
condition $f_1(\omega,t)$ in the OA scheme. In the second part, Section \ref{subsecapprox}, we show, using
the Gaussian as our example, the approximation procedure that allows to treat this frequency distribution within the same framework
employed with the distributions with few poles in the lower half of the complex plane. In Section \ref{secresults} we present the results
that we have obtained by
analyzing the dynamics of the order parameter as derived both from the numerical solution of the equations associated to the OA scheme
and from the numerical integration of the equations of motion of the system of $N$ oscillators, Eqs. (\ref{eq_kura_mod}).
The discussion and the conclusion are in Section \ref{secdiscuss}. The two appendices contain some technical details. In particular,
in Appendix \ref{secappend2} we try to identify the mathematical reason why the essential singularity of the Gaussian frequency
distribution does not have practical consequences on the dynamics of the order parameter when it settles to an
asymptotic nonvanishing value.

\section{The method of approximation}
\label{secmethod}
In this section, we first make a rapid recall of the Kuramoto model and of its treatment by means of the OA ansatz
(although this material can be found in the literature, notably in Ref. \onlinecite{ottant}, where the OA ansatz
has been introduced, we prefer to give here a brief summary, in order to have a self-contained presentation).
Afterwards, we will describe the approximation that we propose in this work, and that is used to study within the OA framework
the case of a Gaussian distribution for the frequencies, in order to circumvent the problem of the essential singularity at
infinity, in the complex plane, of such distribution.

\subsection{The Kuramoto model and the Ott-Antonsen ansatz} 
\label{subseckura} 
In the limit $N \to \infty$, as already anticipated in the Introduction, the dynamics of the system of oscillators
is described by the one-particle distribution function $\rho(\theta,\omega,t)$. The equations of motion
(\ref{eq_kura_mod}) are substituted by the following partial differential equation:
\be
\frac{\partial}{\partial t}\rho(\theta,\omega,t) = -\frac{\partial}{\partial \theta}
\left[\left(\omega + F(\theta,t)\right)\rho(\theta,\omega,t)\right] \, .
\label{contequat}
\ee
This partial differential equation is a continuity equation associated to the conservation of the number of
oscillators with any given frequency \cite{footnote2}. Eq. (\ref{contequat}) conserves the normalization of
$\rho(\theta,\omega,t)$ for any $\omega$, that we choose to set equal to $1$:
\be
\label{normalrho}
\int_0^{2\pi} \dd \theta \, \rho(\theta,\omega,t) = 1 \, .
\ee 
Then, $g(\omega)\rho(\theta,\omega,t)\dd \theta \dd \omega$ is exactly the fraction of oscillators with natural
frequency in the range $(\omega,\omega + \dd \omega)$ that at time $t$ have phases in the range
$(\theta,\theta + \dd \theta)$. In Eq. (\ref{contequat}) $F(\theta,t)$ is the mean field torque, given by
\be
F(\theta,t) = - K\int_{-\infty}^{+\infty} \dd \omega \, \int_0^{2\pi} \dd \theta' g(\omega) \sin (\theta - \theta')
\rho(\theta',\omega,t) \, .
\label{meanforce}
\ee
The complex order parameter $r(t)$ given in Eq. (\ref{deforder_disc}) is now expressed by
\be
r(t) = \int_{-\infty}^{+\infty} \dd \omega \, \int_0^{2\pi} \dd \theta \, g(\omega) \, e^{\ii \theta}
\rho(\theta,\omega,t) \, ,
\label{deforder}
\ee
from which we recall that $|r(t)|\le 1$. The continuity equation (\ref{contequat}) can be expressed in terms
of $r(t)$ as
\be
\frac{\partial}{\partial t}\rho(\theta,\omega,t) = -\frac{\partial}{\partial \theta}
\left[\left(\omega + \frac{K}{2\ii}(r(t)e^{-\ii \theta} - r^*(t)e^{\ii \theta})\right)\rho(\theta,\omega,t)\right] \, ,
\label{fpstart}
\ee
with the star denoting complex conjugation. One can now introduce the Fourier decomposition
\be
\rho(\theta,\omega,t) = \frac{1}{2\pi} \sum_{n=-\infty}^{+\infty} f_n(\omega,t) e^{\ii n\theta} \, .
\label{fourexp}
\ee
Substituting in Eq. (\ref{fpstart}) we transform it in the following systems of partial differential equations for the
Fourier components $f_n(\omega,t)$:
\be
\frac{\partial f_n(\omega,t)}{\partial t} = -\ii n \omega f_n(\omega,t)
-\frac{K}{2}n\left[ r(t) f_{n+1}(\omega,t) - r^*(t) f_{n-1}(\omega,t)\right] \,\,\,\,\,\,\,\,\,\,\,\,\,\,\,\,
n=0,\pm 1,\pm 2,\dots
\label{systemeq}
\ee
Since $\rho(\theta,\omega,t)$ is real and normalized, we have $f_0(\omega,t) \equiv 1$ (and in fact the equation
for $f_0(\omega,t)$ shows that it is constant in time) and $f_{-n}(\omega,t) = f_n^*(\omega,t)$. Besides, the subsystem
for $n<0$ is decoupled from that for $n>0$, so that we need consider only $n>0$. Although not appearing explicitly
in (\ref{systemeq}), the equations for different $\omega$ are globally coupled through the order parameter $r(t)$. The latter
can be obtained as a function of $f_1(\omega,t)$ by plugging the Fourier decomposition (\ref{fourexp}) into the expression
(\ref{deforder}). We have:
\be
r(t)= \int_{-\infty}^{+\infty} \dd \omega \, g(\omega) f_1^*(\omega,t) \, ,
\label{orderfour_1_ap}
\ee

The OA ansatz \cite{ottant} consists in posing
\be
\label{ansoadef}
f_n(\omega,t) = f_1^n(\omega,t) \,\,\,\,\,\,\,\,\,\,\,\,\,\,\,\,\,\, {\rm with} |f_1(\omega,t)|\le 1
\ee
for each $n$; the requirement $|f_1(\omega,t)|\le 1$ allows to have a convergent series in Eq. (\ref{fourexp}).
With this position, it is a simple matter to see that all the equations for positive $n$ are the same, equal to that for $n=1$:
\be
\frac{\partial f(\omega,t)}{\partial t} = -\ii \omega f(\omega,t)
-\frac{K}{2}\left[ r(t) f^2(\omega,t) - r^*(t)\right] \, .
\label{eqneq1}
\ee
where to lighten the notation we have dropped the subscript ``$1$'', since from now on we consider only the first Fourier
component. For convenience of later reference, we rewrite Eq. (\ref{orderfour_1_ap}) with this simplified notation, i.e.,
\be
r(t)= \int_{-\infty}^{+\infty} \dd \omega \, g(\omega) f^*(\omega,t) \, ,
\label{orderfour_1}
\ee
Summing the series in Eq. (\ref{fourexp}) after making the position (\ref{ansoadef}), one easily finds that a positive
function $\rho(\theta,\omega,t)$ is obtained, as it should be for a distribution function.

Obviously the position (\ref{ansoadef}) defines a special class of distribution functions $\rho(\theta,\omega,t)$.
The dynamics determined by the full system (\ref{systemeq}) will in general evolve outside this class. The usefulness of the
position is in the fact that it can be employed to get the asymptotic behavior of the order parameter $r(t)$, as shown below.
The OA ansatz defines a manifold within the space of distribution functions. We see from Eq. (\ref{eqneq1}) that this manifold
is invariant under the dynamics. It is still an infinite dimensional manifold, that we can denote with ${\cal G}_1$.
As stressed in the Introduction, we will consider the dynamics within this
manifold; comments and remarks, concerning in particular the problem of how the dynamics in this manifold is related to the full
dynamics, expressed by Eq. (\ref{fpstart}) or by the system (\ref{systemeq}), are deferred to the concluding Section. We just note
that the stationary solutions of (\ref{fpstart}) belong to the manifold ${\cal G}_1$.
One important point, stressed in Ref. \onlinecite{ottant}, is the following. Let us consider at time $t=0$ a distribution
$\rho(\theta,\omega,0)$ that satisfies the position (\ref{ansoadef}), with the additional properties that the first Fourier
component $f(\omega,0)$ can be analytically continued in the lower half complex $\omega$ plane without presenting there any singularity,
and that $|f(\omega,0)| \to 0$ for ${\rm Im}(\omega) \to -\infty$. Clearly these properties define a submanifold ${\cal G}_2$ of
${\cal G}_1$ in the space of distribution functions. Then, it can be shown \cite{ottant} that if $\rho(\theta,\omega,0)$
belongs to ${\cal G}_2$, then also $\rho(\theta,\omega,t)$ for any $t\ge 0$ belongs to it; i.e., the manifold ${\cal G}_2$
is invariant under the dynamics (\ref{contequat}). We stress that also the manifold ${\cal G}_2$ is infinite dimensional and therefore
it is not yet the low dimensional manifold mentioned above for the order parameter dynamics. This is obtained in the following.

Without assumptions on the frequency distribution $g(\omega)$, the dynamics of our system, within the OA ansatz,
is described by Eqs. (\ref{eqneq1}) and (\ref{orderfour_1}). Then, although we do not have the countable infinity of the index $n$
in (\ref{systemeq}), we still have the continuous infinity of $\omega$, with the equations (\ref{eqneq1}) for different $\omega$
coupled through (\ref{orderfour_1}). As remarked above, one can make a further step in the case
in which $g(\omega)$ can be analytically continued in the complex $\omega$ plane while presenting, as singularities,
only a small number of poles in the lower half plane. This can be seen as follows. Let us first write the complex conjugate
expression of (\ref{orderfour_1}), i.e.,
\be
r^*(t)= \int_{-\infty}^{+\infty} \dd \omega \, g(\omega) f(\omega,t) \, ,
\label{orderfour_2}
\ee
If $g(\omega)$ has the above mentioned properties, and if $f(\omega,t)$ has the properties that make the distribution function
belong to the manifold ${\cal G}_2$, i.e., if $f(\omega,t)$ can be analytically continued without singularities
in the lower half complex $\omega$ plane, with $|f(\omega,t)| \to 0$ for ${\rm Im}(\omega) \to -\infty$, then we can apply the theorem
of residues to Eq. (\ref{orderfour_2}) to obtain \cite{footnote3}:
\be
r^*(t) = - 2\pi \ii \sum_{m=1}^M {\rm Res}\left. \left[g(\omega)f(\omega,t)\right] \right|_{\omega=\omega_m} \, ,
\label{residues}
\ee
where $\omega_1,\dots,\omega_M$ are the poles of $g(\omega)$ in the lower half plane. Since we are assuming that $f(\omega,t)$ has no
singularities in the lower half plane, for each one of the poles of $g(\omega)$ in this half plane we have
\be
{\rm Res}\left. \left[g(\omega)f(\omega,t)\right] \right|_{\omega=\omega_m} = f(\omega_m,t)
{\rm Res}\left. \left[g(\omega)\right] \right|_{\omega=\omega_m} \, ,
\label{residues_s}
\ee
Defining now the quantities $a_m$ by
\be
a_m = - 2\pi \ii {\rm Res}\left. \left[g(\omega)\right] \right|_{\omega=\omega_m} \,\,\,\,\,\,\,\,\,\,\,\,\,\,\,\,\,
m=1,\dots,M,
\label{defam}
\ee
and the complex dynamical variables $r^{(m)}(t)$ by
\be
\label{rpart}
r^{(m)*}(t) = f(\omega_m,t) \,\,\,\,\,\,\,\,\,\,\,\,\,\,\,\,\, m=1,\dots,M,
\ee
Eq. (\ref{residues}) can be written as
\be
r^*(t) = \sum_{m=1}^M a_m r^{(m)*}(t) \, .
\label{residuesb}
\ee
From Eqs. (\ref{eqneq1}) and (\ref{rpart}) one obtains the equations for the time evolution of the variables $r^{(m)}(t)$, i.e.
\be
\frac{\dd r^{(m)*}(t)}{\dd t} = -\ii \omega_m r^{(m)*}(t) - \frac{K}{2} \left[ r(t) [r^{(m)*}(t)]^2 - r^*(t) \right]
\,\,\,\,\,\,\,\,\,\,\,\,\,\,\,\,\, m=1,\dots,M,
\label{eqforrst}
\ee
with $r(t)$ given by (\ref{residuesb}). We will sometimes refer also to the complex conjugate of Eqs. (\ref{eqforrst}), i.e.,
\be 
\frac{\dd r^{(m)}(t)}{\dd t} = \ii \omega_m^* r^{(m)}(t) - \frac{K}{2} \left[ r^*(t) r^{(m)2}(t) - r(t) \right]
\,\,\,\,\,\,\,\,\,\,\,\,\,\,\,\,\, m=1,\dots,M.
\label{eqforrd}
\ee
These equations, together with (\ref{residuesb}), constitute a system of $M$ coupled complex ordinary differential equations, determining
the dynamics of the order parameter $r(t)$. Of course, this system contains less information than Eq. (\ref{eqneq1}) (where, loosely
speaking, instead of the discrete index $m$ taking only $M$ values we have a continuous infinity of indices represented by the frequency
$\omega$), but, under the hypotheses assumed for the derivation of the system, it represents exactly the dynamics of the order parameter.
We have therefore arrived at the low dimensional manifold, that we can denote by ${\cal M}$, in which the dynamics of the order parameter evolves;
it is the manifold spanned by the dynamical variables $r^{(m)}$. We repeat that comments of the relation between the full dynamics of
the system and the order parameter dynamics determined by the system (\ref{eqforrst}) are found in the discussion section.

Summarizing: {\it i)} the infinite dimensional manifold ${\cal G}_1$ defined by the OA ansatz is invariant under the dynamics
(\ref{contequat}); {\it ii)} the infinite dimensional manifold ${\cal G}_2$, submanifold of ${\cal G}_1$, is also invariant;
{\it iii)} if the frequency distribution $g(\omega)$ has only poles in the lower half of the complex $\omega$ plane {\it and} the
distribution function at time $0$ belongs to the manifold ${\cal G}_2$, there is a low dimensional manifold
${\cal M}$ in which the order parameter dynamics evolves.

\subsubsection{The role of the initial condition $f(\omega,0)$}
\label{secinitial}
The main purpose of this paper is to study the dynamics of the order parameter in the case in which the frequency distribution $g(\omega)$
cannot be analytically continued to the lower half complex $\omega$ plane while presenting there only (few) poles. However, we have seen
that an important ingredient to obtain the system of $M$ equations (\ref{eqforrst}) is the assumption about the initial value $f(\omega,0)$,
allowing the distribution function at time $t=0$ to belong to the manifold ${\cal G}_2$. In fact, as we have seen in the
previous section, to obtain the system (\ref{eqforrst}) for the description of the dynamics of the order parameter when $g(\omega)$ has
only poles in the lower half plane, one needs also the assumption that $f(\omega,0)$ can be analytically continued in the lower half plane
without singularities. Thus, before addressing our main problem, in this section we would like to consider the practical relevance of
this assumption about the initial value $f(\omega,0)$ for the evaluation of the long-time behavior of the order parameter.

As described with more details in Section \ref{secresults}, where the results of our simulations are presented, the dynamics
of the order parameter has been computed in three ways, making a comparison of the long-time behavior so obtained: from the numerical
integration of the $N$-body equations of motion (\ref{eq_kura_mod}), from the numerical solution of the partial differential equation
(\ref{eqneq1}), supplemented with
Eq. (\ref{orderfour_1}), for the Fourier component $f(\omega,t)$, and from the numerical solution of the system of ordinary differential
equations (\ref{eqforrst}). While details on the numerical integrations and on their comparison are given in Section \ref{secresults}, here we
anticipate an interesting result, devoting
the remaining of this section to provide a justification for it. To perform the numerical integration of the partial differential equation
(\ref{eqneq1}) one has to start from an initial condition $f(\omega,0)$. We have performed the integration choosing two different initial conditions
(in both cases given by differentiable functions on the real line),
one in which the function $f(\omega,0)$ can be analytically continued in the lower half plane without singularities, tending to $0$ for
${\rm Im}(\omega) \to -\infty$, and another one in which $f(\omega,0)$ can be continued to the lower half plane but having there a singularity. The comparison of the order parameter dynamics obtained by these simulations with the one derived from the integration of the equations
(\ref{eqforrst})
has shown that: when $f(\omega,0)$ can be analytically continued without singularities, the dynamics of the order parameter obtained from
Eq. (\ref{eqneq1}) coincides with that derived from the equations (\ref{eqforrst}) (obviously with the same initial value of the order
parameter); this is expected from the procedure used to obtain these latter
equations. On the other hand, when $f(\omega,0)$ has singularities in the lower half plane, the dynamics of the order parameter obtained from
Eq. (\ref{eqneq1}) differs from that derived from the equations (\ref{eqforrst}), but, and this is the important point, the long-time
behaviors coincide. As said above, we want to give a justification for this.

To this aim, in this section we consider the case in which
$g(\omega)$ can be analytically continued to the lower half complex $\omega$ plane having there only poles.
As a first thing we stress that the analysis in this section will not be mathematically rigorous, but hopefully could be sufficient
to convince the reader that the assumptions about $f(\omega,0)$, although mathematically necessary to obtain the system of equations
(\ref{eqforrst}) (or(\ref{eqforrd})), are not physically relevant for the computation of the asymptotic states of the order parameter.
In this spirit, let us consider the linearization of Eq. (\ref{eqneq1}), i.e., the equation without the term proportional to
$r(t)f^2(\omega,t)$, which is of third order in $|f(\omega,t)|$. This means that we study the dynamics determined by Eq. (\ref{eqneq1})
when $|f(\omega,t)|$ is sufficiently small, to see the relation of this dynamics to that determined, in the same limit, by the set of
equations (\ref{eqforrst}). We then study the partial differential equation
\be
\label{eqneq1lin}
\frac{\partial f(\omega,t)}{\partial t} +\ii \omega f(\omega,t) -\frac{K}{2} r^*(t) = 0 \, ,
\ee
where we remind that 
\be
r^*(t)= \int_{-\infty}^{+\infty} \dd \omega \, g(\omega) f(\omega,t) \, .
\label{orderfour_lin}
\ee
We want to analyze the relation of the dynamics of $r^*(t)$ as determined by (\ref{eqneq1lin}) and (\ref{orderfour_lin}) with that
determined by the linear limit of Eqs. (\ref{eqforrst}), i.e.,
\be
\frac{\dd r^{(m)*}(t)}{\dd t} = -\ii \omega_m r^{(m)*}(t) + \frac{K}{2} r^*(t)
\,\,\,\,\,\,\,\,\,\,\,\,\,\,\,\,\, m=1,\dots,M,
\label{eqforrst_lin}
\ee
with $r^*(t)$ given by Eq. (\ref{residuesb}).
It is interesting to note that the structure of Eq. (\ref{eqneq1lin}) is that of the conjugate of a linearized Vlasov equation
for the Fourier components of a distribution function of the position $\theta$ and of the fictitious momentum $\omega$;
it can be solved with the following procedure \cite{footnote4}.
To study the initial value
problem associated to Eq. (\ref{eqneq1lin}) we use the Laplace transform in time, that defines $\widetilde{f}(\omega,\nu)$ by
\be
\label{laplforf}
\widetilde{f}(\omega,\nu) = \int_0^\infty \dd t \, e^{\ii \nu t} f(\omega,t) \, ,
\ee
where, as usual with this transform, ${\rm Im}(\nu)$ must be positive and sufficiently large to assure the existence of the
integral \cite{footnote5}. The inversion formula is given by
\be
\label{invlap}
f(\omega,t) = \frac{1}{2\pi} \int_{{\cal C}} \dd \nu \, e^{-\ii \nu t} \widetilde{f}(\omega,\nu) \, ,
\ee
where the integration path ${\cal C}$ must pass above all the singularities (e.g., with a line parallel to the real axis) of the function
$\widetilde{f}(\omega,\nu)$. Taking the Laplace transform of Eq. (\ref{eqneq1lin}) and inserting the explicit expression of $r^*(t)$
given by (\ref{orderfour_lin}) we have
\be
\label{eqneq1lap}
-f(\omega,0) - \ii \nu \widetilde{f}(\omega,\nu) + \ii \omega \widetilde{f}(\omega,\nu) -\frac{K}{2}
\int \dd \omega' \, \widetilde{f}(\omega',\nu) g(\omega') = 0 \, ,
\ee
where $f(\omega,0)$ is the initial value, at time $t=0$, of $f(\omega,t)$. For convenience we have not explicitly indicated
the integration limits $(-\infty,+\infty)$ in the integral over $\omega'$; we will do the same in the integrals over $\omega$
appearing in the remaining of this section. The last equation can be rewritten as
\be
\label{eqneq1lap1}
\widetilde{f}(\omega,\nu) - \ii \frac{K}{2}\frac{1}{\nu-\omega}
\int \dd \omega' \, \widetilde{f}(\omega',\nu) g(\omega') = \frac{f(\omega,0)}{\ii(\omega-\nu)} \, .
\ee
Multiplying by $g(\omega)$ and integrating over $\omega$ one obtains
\be
\label{eqforint}
\int \dd \omega' \, \widetilde{f}(\omega',\nu) g(\omega') = \frac{1}{\epsilon(\nu)}
\int \dd \omega \, \frac{f(\omega,0)g(\omega)}{\ii(\omega-\nu)} \, ,
\ee
where the function $\epsilon(\nu)$ is defined by
\be
\label{defindiel}
\epsilon(\nu) = 1 - \ii \frac{K}{2} \int \dd \omega \, \frac{g(\omega)}{\nu-\omega} \, .
\ee
It must be stressed that the integral in the right hand side defines the function $\epsilon(\nu)$ for ${\rm Im}(\nu) >0$;
for the remaining of the complex $\nu$ plane, $\epsilon(\nu)$ is defined by the analytical continuation of the
right hand side of (\ref{defindiel}). Using Eq. (\ref{eqforint}) in the integral appearing in
Eq. (\ref{eqneq1lap}), and multiplying by $g(\omega)$, one finds
\be
\label{eqneq1lap2}
g(\omega)\widetilde{f}(\omega,\nu) = \frac{f(\omega,0)g(\omega)}{\ii (\omega -\nu)} + \frac{1}{\epsilon(\nu)}\frac{K}{2}
\frac{g(\omega)}{\ii (\omega -\nu)}\int \dd \omega' \, \frac{f(\omega',0)g(\omega')}{\ii (\omega' -\nu)} \, .
\ee
The integral over $\omega$ of the left hand side gives, according to Eq. (\ref{orderfour_lin}), the Laplace transform of
$r^*(t)$, i.e.,
\be
\label{laptransorder}
\widetilde{r}^*(\nu) = \int_0^\infty \dd t \, e^{\ii \nu t} r^*(t) =
\int \dd \omega \, g(\omega) \widetilde{f}(\omega,\nu) \, .
\ee
Substituting the right hand side of Eq. (\ref{eqneq1lap2}), and using Eq. (\ref{defindiel}), we obtain
\be
\label{laptransorder_1}
\widetilde{r}^*(\nu) = \frac{1}{\epsilon(\nu)}
\int \dd \omega \, \frac{f(\omega,0)g(\omega)}{\ii (\omega -\nu)} \, .
\ee
In this way $\widetilde{r}^*(\nu)$ is obtained as a function of the frequency distribution
$g(\omega)$ and of the initial value $f(\omega,0)$. By inversion one can then get $r^*(t)$:
\be
\label{invtogetr}
r^*(t) = \frac{1}{2\pi} \int_{{\cal C}} \dd \nu \, e^{-\ii \nu t} \widetilde{r}^*(\nu) \, .
\ee

To make a comparison with the $r(t)$ determined by the set of equations (\ref{eqforrst_lin}) we consider the simple case in
which $M=1$. Since in this case $r^*(t)$ is simply proportional to the factor $a_1$ in Eq. (\ref{residuesb}), we have
\be
\frac{\dd r^*(t)}{\dd t} = -\ii \omega_1 r^*(t) + \frac{K}{2} r^*(t) \, .
\label{eqforrst_lin_1}
\ee
We have $M=1$ when, for example, $g(\omega)$ is a Lorentzian:
\be
\label{gloren}
g(\omega) = \frac{b}{\pi}\frac{1}{\omega^2 + b^2} \, ,
\ee
with $b$ a given positive number. In this case we have $\omega_1 = -\ii b$, and Eq. (\ref{eqforrst_lin_1}) becomes
\be
\frac{\dd r^*(t)}{\dd t} = - b r^*(t) + \frac{K}{2} r^*(t) \, .
\label{eqforrst_lin_2}
\ee
The solution of this simple equation is clearly \cite{footnote6}
\be
\label{sol_rst_lin_2}
r^*(t) = r^*(0) e^{\left(\frac{K}{2} - b\right) t} \, .
\ee
We have to compare this solution with that obtained from Eq. (\ref{eqneq1lap2}) when $g(\omega)$ is given by (\ref{gloren}).
Looking at Eq. (\ref{eqneq1lap2}) and considering Eq. (\ref{defindiel}), we see that to compute the integral over $\omega$ of the right
hand side of (\ref{eqneq1lap2}) we need to evaluate the two following integrals:
\be
\label{neededint}
\int \dd \omega \, \frac{f(\omega,0)g(\omega)}{\omega-\nu} \, ,\,\,\,\,\,\,\,\,\,\,\,\,\,\,\,\,\,\,\,\,\,\,\,\,\,\,\,\,\,
\int \dd \omega \, \frac{g(\omega)}{\omega -\nu}
\ee
for ${\rm Im}(\nu)>0$. Some details of the computations presented in the remaining of this section are given in Appendix \ref{secappend1}.
We will consider two cases: one in which $f(\omega,0)$ has the properties required by the definition of the manifold ${\cal G}_2$, and
another in which $f(\omega,0)$ does not have them. The second integral in Eq. (\ref{neededint}) does not depend on $f(\omega,0)$ and
therefore its evaluation is the same in both cases. With $g(\omega)$ as in Eq. (\ref{gloren}) we find
\be
\label{value_int2}
\int \dd \omega \, \frac{g(\omega)}{\omega -\nu} = -\frac{1}{\nu + \ii b} \, .
\ee
From this we also obtain the function $\epsilon(\nu)$, i.e.,
\be
\label{specdiel}
\epsilon(\nu) = \frac{\nu -\ii \left( \frac{K}{2} - b \right)}{\nu + \ii b} \, .
\ee

Let us begin with the case in which $f(\omega,0)$ satisfies the properties related to the manifold ${\cal G}_2$; then, the first integral
in Eq. (\ref{neededint}) is obtained as
\be
\label{value_int1_a}
\int \dd \omega \, \frac{f(\omega,0)g(\omega)}{\omega-\nu} = - \frac{f(-\ii b,0)}{\nu + \ii b} \, .
\ee
Using the last three equations in (\ref{eqneq1lap2}) and (\ref{laptransorder}) we find
\be
\label{rnu_1}
\widetilde{r}^*(\nu) = \frac{\ii f(-\ii b,0)}{\nu - \ii \left( \frac{K}{2} - b \right)} \, .
\ee
Then, the inverse Laplace transform allows to have
\be
\label{roft_1}
r^*(t) = f(-\ii b,0) e^{\left(\frac{K}{2} - b\right) t} \, .
\ee
Furthermore, from Eq. (\ref{orderfour_lin}) we get that $r^*(0) = f(-\ii b,0)$, implying that, as we expect, Eq. (\ref{roft_1}) is identical
to Eq. (\ref{sol_rst_lin_2}).

We now consider the case in which $f(\omega,0)$ has singularities in the lower half complex $\omega$ plane. To have a definite example,
we take also for $f(\omega,0)$ a Lorentzian form:
\be
\label{fomega0def}
f(\omega,0) = \frac{c}{\pi}\frac{1}{\omega^2 + c^2} \, ,
\ee
with $c$ a positive number different from $b$. In this case the initial value $r^*(0)$, using Eq. (\ref{orderfour_lin}), is given by
\be
\label{rteq0lor}
r^*(0) = \frac{1}{\pi(b+c)} \, .
\ee
On the other hand, now the first integral in (\ref{neededint}) is evaluated to be
\be
\label{value_int1_b}
\int \dd \omega \, \frac{f(\omega,0)g(\omega)}{\omega-\nu} = -\frac{1}{\pi(b^2-c^2)}
\frac{(b-c)\nu + \ii (b^2 -c^2 )}{(\nu + \ii b)(\nu + \ii c)} \, .
\ee
Plugging in Eq. (\ref{eqneq1lap2}) and using Eq. (\ref{defindiel}), from Eq. (\ref{laptransorder}) we have
\be
\label{rnu_2}
\widetilde{r}^*(\nu) = \frac{1}{\pi(b^2-c^2)}
\frac{\ii (b-c)\nu - (b^2 -c^2 )}{\left[\nu - \ii \left( \frac{K}{2} - b \right)\right]\left[\nu + \ii c\right]} \, .
\ee
By inverse Laplace transform we get, using (\ref{rteq0lor}),
\be
\label{roft_2}
r^*(t) = \frac{r^*(0)}{\frac{K}{2} + c -b}\left[ \left( \frac{K}{2}+c\right) e^{\left(\frac{K}{2}-b\right)t} -b e^{-ct}\right]
\, .
\ee
Then, we see that now the solution of Eq. (\ref{eqneq1lin}) does not coincide with that of Eq. (\ref{eqforrst_lin_2});
therefore the latter does not represent exactly the dynamics of the former.

But now we invoke physical considerations, leaving aside mathematical rigor, to argue that the difference between
Eq. (\ref{roft_2}) and Eq. (\ref{sol_rst_lin_2}) is not physically relevant, in general, for the evaluation of the asymptotic states
of the order parameter. In fact, we note that the second
term in (\ref{roft_2}) vanishes exponentially. Then, if also the first term does so, when $\left( \frac{K}{2} - b\right) <0$,
we have in any case a dynamics that leads to an exponentially decaying $r^*(t)$.
On the other hand, if $\left( \frac{K}{2} - b\right) >0$
and the first term in (\ref{sol_rst_lin_2}) increases exponentially, the second term will give a negligible contribution. In this latter
case, the nonlinear term of Eq. (\ref{eqneq1}) will come into play to lead $r^*(t)$ to its asymtpotic value.

As emphasized at the beginning of this section, this analysis is not rigorous, since it is based on the linearization
of Eqs. (\ref{eqneq1}) and (\ref{eqforrst}); the ultimate test should be given by the comparison of the numerical solution of the full equations
(\ref{eqneq1}) and (\ref{eqforrst}). This is done below in Section \ref{secresults}; there we show that while the solution of Eq. (\ref{eqneq1})
and (\ref{eqforrst}) produce the same dynamics of the order parameter when $f(\omega,0)$ does not have singularities in the lower half plane,
the two dynamics differ initially when $f(\omega,0)$ does have singularities in the lower half plane, but nevertheless the asymptotic
behavior of the order parameter is the same, and this is the important point for the determination of the phase diagram.
In conclusion, we can expect that when the frequency distribution $g(\omega)$ has only poles in the lower half plane, the system of
ordinary differential equations equations (\ref{eqforrst}) (or (\ref{eqforrd})) can be taken as a substitute, {\it as far as the
long-time evolution of the order parameter is concerned}, for the description of the dynamics of Eq. (\ref{eqneq1})
independently from the presence of singularities, in the lower half of the complex $\omega$ plane, of the analytic continuation
of the initial value $f(\omega,0)$.
When $f(\omega,0)$ does not have the properties required by the definition of the manifold ${\cal G}_2$, the two dynamics will present an
initial transient in which they differ somewhat, but afterwards they will reach the same asymptotic value of the order parameter, as
shown in Section \ref{secresults}.

The above analysis has been done for frequency distributions $g(\omega)$ that can be analytically continued in the lower half plane without
singularities. It can be argued that the same conclusion about the role of the initial condition $f(\omega,0)$ holds for generic
$g(\omega)$, since also in that case singularities of $f(\omega,0)$ in the lower half plane will contribute exponentially decaying terms
in the dynamics of the order parameter. In the section of the results we provide an example concerning this point.
 
\subsection{The case of a Gaussian frequency distribution}
\label{subsecapprox}

We now come to the main part of this work, in which we consider the case of a frequency distribution $g(\omega)$ that cannot be
analytically continued to complex values presenting only poles in the lower half plane and tending to zero for ${\rm Im}(\omega) \to -\infty$.
We take the physically relevant example of a Gaussian frequency distribution.
If we want to study the case in which the frequency distribution $g(\omega)$ is a Gaussian, the scheme described above cannot be
employed, since this function, when considered in the complex $\omega$ plane, has an essential singularity at infinite, and therefore
the application of the residue theorem as in Eq. (\ref{residues}) is not possible. As emphasized in the introductory remarks, on physical
grounds one expect that frequency distributions that numerically do not differ very much should lead to very similar dynamics for the order
parameter $r(t)$. After all, even if two numerically similar frequency distributions (for real $\omega$) have very different mathematical
properties concerning their analytical continuation on the complex $\omega$ plane, we do not expect this to have a strong influence at
least on the most relevant and visible properties of the dynamics of the order parameter.

\begin{figure}[ht]
\begin{center}
\includegraphics[scale=0.5,trim= 0cm 0cm 0cm 0cm,clip]{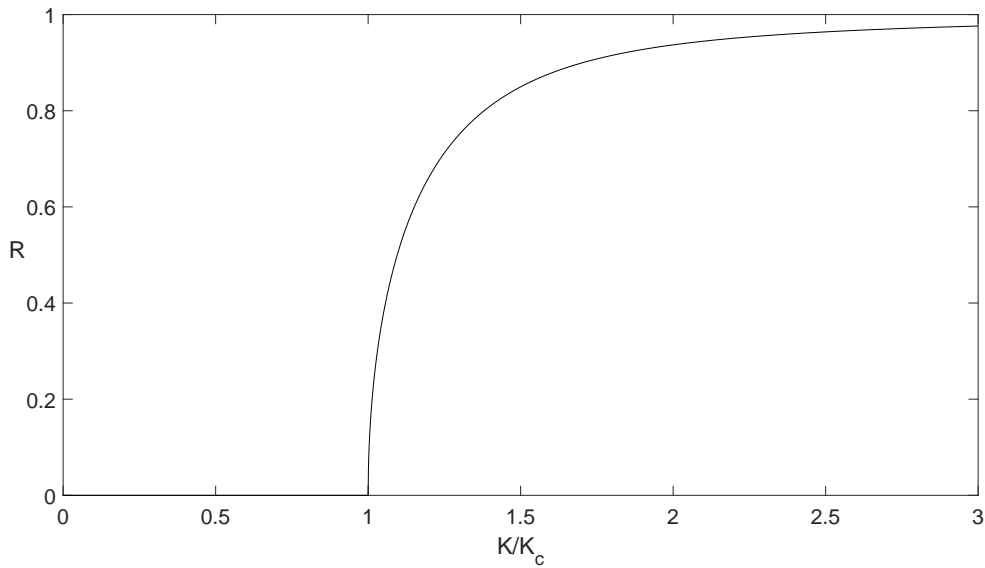}
\caption{The function $R(K/K_c)$ in the case of the Gaussian distribution (\ref{gaussfreq}), computed from Eq. (\ref{rofkesp}).
The critical value of the coupling constant is $K_c = 2\sqrt{2}\sigma/\sqrt{\pi}$.}
\label{figrofk}
\end{center}
\end{figure}

We take as working example a symmetric unimodal distribution, in particular a Gaussian, in order to have the simplest framework in which to
describe our approach. It will be clear that the approach could be applied in other cases, like e.g., bimodal distributions, that give rise
to more elaborate and interesting phase diagrams \cite{bimodal}. Let us then take the frequency distribution
\be
\label{gaussfreq}
g(\omega) = \frac{1}{\sqrt{2\pi \sigma^2}} e^{-\frac{\omega^2}{2\sigma^2}} \, .
\ee
As anticipated before, symmetric unimodal frequency distributions present a rather simple $R(K)$, describing the degree of synchronization
of the system as a function of the coupling constant $K$. Following the Kuramoto analysis, the system of oscillators with this frequency
distribution goes to an incoherent asymptotic state, $R=0$, if the coupling constant $K$ is not larger than the threshold value
$K_c = 2/(\pi g(0)) = 2\sqrt{2}\sigma/\sqrt{\pi}$; above this threshold, the system is partially synchronized, $R(K)$ bifurcates continuously
at $K=K_c$ from $R=0$ and grows monotonically \cite{strogatz2000}, tending to $1$ for $K \to \infty$. For comparison with the results
presented in the following, we evaluate $R(K)$ in our case. The function $R(K)$ for $K\ge K_c$ for a given unimodal symmetric
$g(\omega)$ is given implicitly by \cite{strogatz2000,briefbook}:
\be
\label{rofk}
K \int_{-\frac{\pi}{2}}^{\frac{\pi}{2}} \dd \theta \, \cos^2 \theta \, g(KR\sin \theta) = 1 \, .
\ee
When $g$ is given by Eq. (\ref{gaussfreq}) we obtain in few passages the following expression
\be
\label{rofkesp}
\frac{K}{2\sigma}\sqrt{\frac{\pi}{2}}\, e^{-\frac{K^2R^2}{4\sigma^2}} \left[ I_0 \left(\frac{K^2R^2}{4\sigma^2}\right) +
I_1 \left(\frac{K^2R^2}{4\sigma^2}\right)\right] = 1 \, ,
\ee
where $I_0$ and $I_1$ are the modified Bessel functions of order $0$ and order $1$, respectively.
In Fig. \ref{figrofk} we plot the function $R(K/K_c)$ obtained by numerically solving this equation. For completeness the plot starts at $K=0$,
to show that for $K\le K_c$ the value of $R$ vanishes. As it could have been guessed on physical grounds,
Eq. (\ref{rofkesp}) shows that $R$ depends on the ratio $K/\sigma$. Therefore, without loss of generality, and in order to simplify the
notations, we can work at the given value $\sigma= 1/\sqrt{2}$, so that our Gaussian distribution will be
\be
\label{gaussfreqnorm}
g(\omega) = \frac{1}{\sqrt{\pi}} e^{-\omega^2} \, .
\ee
Correspondingly, the threshold value of the coupling constant will be $K_c = 2/\sqrt{\pi}$.

Let us now proceed with our approximation, which is actually quite simple. It consists in writing $e^{-\omega^2}=1/e^{\omega^2}$, and
then replacing the exponential in the denominator with its power expansion truncated at a given order. Namely, we replace $g(\omega)$
with $g_P(\omega)$ given by
\be
\label{apprgauss}
g_P(\omega) = \frac{1}{C_P\sqrt{\pi}} \left[ \sum_{s=0}^P \frac{1}{s!} \omega^{2s} \right]^{-1} \, .
\ee  
The coefficient $C_P$ is necessary since we must work with normalized
frequency distributions, and with a finite $P$ the integral of the fraction substituting $e^{-\omega^2}$ is slightly larger than $\sqrt{\pi}$.
Therefore, we already see that the approximation implies a change in the threshold value, that becomes $K_c^{(P)} = 2 C_P/\sqrt{\pi}$.
However, we will see that already for quite small values of $P$, the value of $C_P$ is so close to $1$ that the threshold value $K_c$
becomes practically coincident to that of the exact Gaussian distribution. Actually, the whole curve $R_P(K)$ will be extremely
close to the exact one. As a first hint of this, before explicitly computing $R_P(K)$, we can take the following fact. We
know \cite{strogatz2000,briefbook} that from Eq. (\ref{rofk}) one can obtain an expansion of $R^2(K)$ in terms of the powers of
$(K-K_c)$, in which the coefficients depend on the derivatives of $g(\omega)$ at $\omega=0$, with derivatives of increasing order
coming into play at increasing powers of $K-K_c$. Now, we can show that the first $2P$ derivatives of $g_P(\omega)/C_P$ at $\omega=0$
coincide with those of the Gaussian (\ref{gaussfreq}). Thus, apart from the discrepancy due to the factor $C_P$,
that as anticipated is extremely close to $1$ already for small values of $P$, this indicates that the curve $R_P(K)$ should be very
close to the exact $R(K)$. The equality just mentioned of the first $2P$ derivatives at $\omega=0$ can be proved as in the following.
Obviously we have to consider only the derivatives of even order, since $g(\omega)$ and $g_P(\omega)$ are even functions, and then the
odd derivatives at $\omega=0$ vanish. It is equivalent to work with the coefficients of the power expansions, proportional to the
derivatives. Neglecting also the multiplicative coefficient $1/\sqrt{\pi}$, we start with the trivial power expansion
\be
\label{expangauss}
e^{-\omega^2} = \sum_{k=0}^{\infty} \frac{(-1)^k}{k!} \omega^{2k}
\ee
(thus, the $(2k)$-th derivative at $\omega=0$ is equal to $(-1)^k (2k)!/k!$). Concerning $g_P(\omega)$, we write the power expansion as
\be
\label{expangr}
\left[ \sum_{s=0}^P \frac{1}{s!} \omega^{2s} \right]^{-1}
= \sum_{k=0}^P c_k \omega^{2k} + \sum_{k=P+1}^{\infty} d_k \omega^{2k} \, ,
\ee
where we have separated the first terms up to the power $2P$ from the remaining ones. We are interested in the coefficients
$c_k$, related to the first $2P$ derivatives at $\omega=0$ of the left hand side. Taking $\omega=0$ one sees that $c_0 = 1$.
We rewrite the above expression as
\be
\label{expangrb}
\left[ \sum_{s=0}^P \frac{1}{s!} \omega^{2s} \right]
\left[ \sum_{k=0}^P c_k \omega^{2k} + \sum_{k=P+1}^{\infty} d_k \omega^{2k} \right] = 1 \, .
\ee
Performing the multiplication one can obtain linear equations in the unknown coefficients $c_k$ and $d_k$. It is easy to see
that the coefficients $c_1,\dots,c_P$ are obtained by the following system of $P$ linear equations:
\be
\label{syscr}
\sum_{k=0}^q \frac{1}{k!} c_{q-k} = 0 \,\,\,\,\,\,\,\,\,\,\,\,\,\,\,\,\,\,\, q=1,2,\dots,P \, .
\ee
Note that the system is not homogeneous, since in each equation the last term of the sum is proportional to the known
coefficient $c_0=1$. It is immediate to see that taking $c_k = \frac{(-1)^k}{k!}$ this system is satisfied, since for any
$q$ the sum would be equal to $q!$ times the binomial expansion of $(1-1)^q = 0^q = 0$. Thus, from the uniqueness of the coefficients
of a power expansion, we find that the first coefficients, up to $k=2P$, of the expansion of the left hand
side of Eq. (\ref{expangr}) are equal to those of the expansion of the Gaussian (\ref{expangauss}).

\section{Comparison of the results for the order parameter dynamics}
\label{secresults}
\begin{figure}[ht]
\begin{center}
\includegraphics[scale=0.5,trim= 0cm 0cm 0cm 0cm,clip]{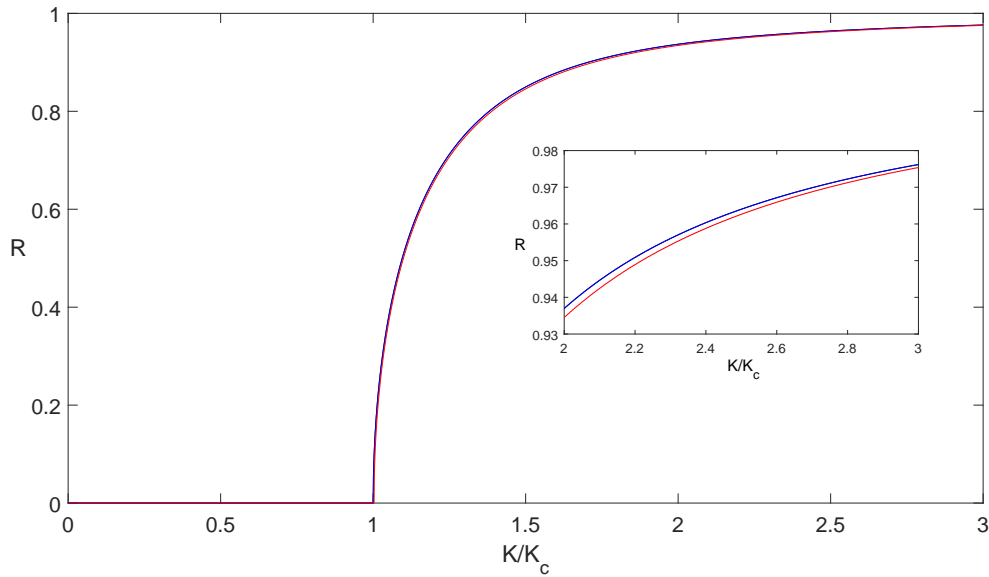}
\caption{The functions $R(K/K_c)$ for the Gaussian distribution (\ref{gaussfreqnorm}) (in black), for the distribution (\ref{apprgauss}) with
$P=12$ (in blue) and with $P=6$ (in red). The inset shows the large $K/K_c$ range. Actually the blue curve, in the scale of the plot, even in the inset,
superimposes to the black curve, and the two curves cannot be distinguished.}
\label{figrofkcomp}
\end{center}
\end{figure}

We begin by providing the following result concerning the value of the parameter $C_P$, related to the normalization of the sum in
Eq. (\ref{apprgauss}), for two chosen values of $P$. For $P=6$ we have $C_6 \approx 1.003$, while for $P=12$ we have $C_{12}\approx 1.00003$.
Therefore, we have $K_c^{(6)} \approx 1.003 K_c$ and $K_c^{(12)} \approx 1.00003 K_c$. This implies that these threshold values, especially
the one for $P=12$, are extremely close to that of the exact Gaussian distribution. In Fig. \ref{figrofkcomp} we plot, as a function of $K/K_c$,
the modulus $R(K/K_c)$ of the order parameter for the Gaussian (\ref{gaussfreqnorm}), $g(\omega)$, the modulus $R_6(K/K_c)$ for the distribution
(\ref{apprgauss}) with $P=6$, $g_6(\omega)$, and the modulus $R_{12}(K/K_c)$ for the distribution (\ref{apprgauss}) with $P=12$,
$g_{12}(\omega)$.
Actually, the whole curve $R_{12}(K/K_c)$ is so close to $R(K/K_c)$ that the two curves cannot be
distinguished, in the scale of the plot. This is true also for the inset, where we plot the curves for large values of $K/K_c$. Concerning
the curve $R_6(K/K_c)$, it can hardly be distinguished from $R(K/K_c)$ in the main plot, and only in the inset the very small difference
can be clearly seen.

\begin{figure}[ht]
\begin{center}
\includegraphics[scale=0.5,trim= 0cm 0cm 0cm 0cm,clip]{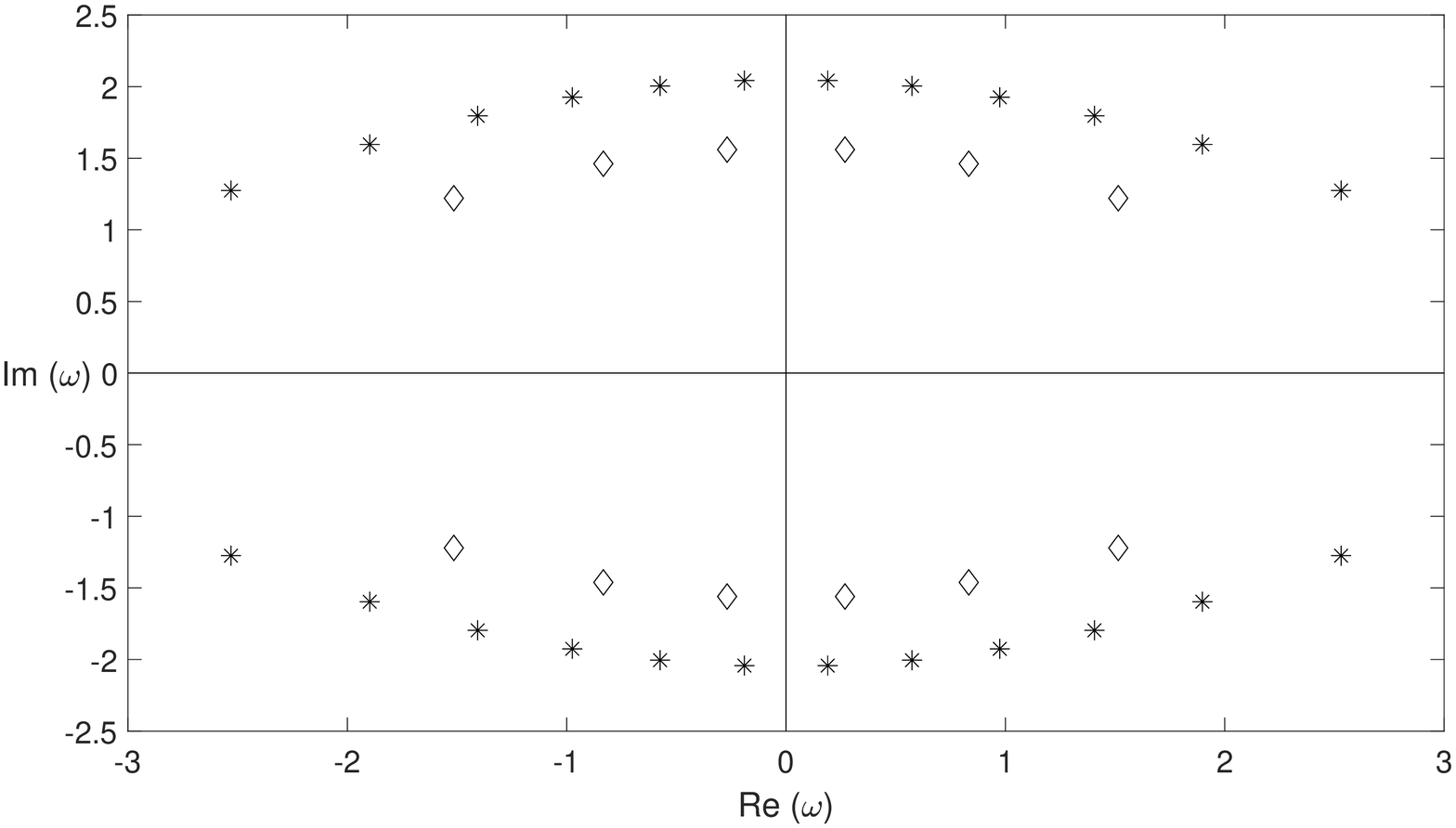}
\caption{The $12$ poles of the function $g_6(\omega)$ (diamonds) and the $24$ poles of the function $g_{12}(\omega)$ (stars) in the
complex $\omega$ plane.}
\label{figpoli}
\end{center}
\end{figure}

The sum in Eq. (\ref{apprgauss}) has $2P$ zeros in the complex $\omega$ plane, i.e., the function $g_P(\omega)$ has $2P$ poles; obviously,
they are all complex, since the coefficients in the sum are all positive and the powers of $\omega$ are even. It is easy to see that, if
$\omega_m$ is a pole with
${\rm Re}(\omega_m) \ne 0$, then also $\omega_m^*$, $-\omega_m$ and $-\omega_m^*$ are poles.
In Fig. \ref{figpoli} we show the poles of $g_6(\omega)$ and of $g_{12}(\omega)$ in the complex $\omega$ plane. The poles in the lower half plane are those
that appear in Eqs. (\ref{eqforrst}) and (\ref{eqforrd}).

We now proceed to compare the results of the dynamics. We will use the approximation with $P=12$. In Figs. \ref{figdynam_1} and
\ref{figdynam_2}, see below, we will plot the time course of the modulus $R(t)$ of the complex order parameter as obtained in three ways
(as anticipated in Section \ref{secinitial}): from the numerical integration of the $N$-body equations of motion (\ref{eq_kura_mod}), from
the numerical solution of the partial differential equation (\ref{eqneq1}), supplemented with Eq. (\ref{orderfour_1}), for the Fourier
component $f(\omega,t)$, and from the numerical solution of the system of ordinary differential equations (\ref{eqforrst}) (or equivalently
(\ref{eqforrd})). It is important to emphasize that, in the spirit of this work, Eq. (\ref{eqneq1}) has been solved using the Gaussian
distribution function (\ref{gaussfreqnorm}); this is done in order to compare the results of the dynamics employing the exact frequency
distribution, for which one cannot use the system of few equations (\ref{eqforrst}) for the variables $r^{(m)*}(t)$, and that of the dynamics
where the system can be used, but at the price of approximating the frequency distribution. For all the numerical integrations
one has to choose the initial conditions. The remaining of this Section is divided in two parts. In Section \ref{secresults1} we provide the
details on the choice of initial conditions, while in Section \ref{secresults2} we present and comment the results of the simulations.

\subsection{The choice of the initial conditions}
\label{secresults1}
Let us begin with the description of the initial condition $f(\omega,0)$ for the numerical integration of the partial differential equation
(\ref{eqneq1}). As already mentioned in Section \ref{secinitial}, we have performed this integration choosing two different initial conditions
for $f(\omega,0)$, given respectively by
\be
\label{finit1}
f(\omega,0) = \frac{1}{5} e^{-\ii \omega}
\ee
and
\be
\label{finit2}
f(\omega,0) = \frac{\delta}{\omega^2 +1} \, .
\ee
In (\ref{finit1}) the coefficient $1/5$ has been arbitrarily chosen (we remind that the OA scheme requires $|f(\omega,0)|\le 1$, see
Eq. (\ref{ansoadef})). That functional form of $f(\omega,0)$ can be analytically continued without singularities in the lower half plane,
with $|f(\omega,0)|\to 0$ for ${\rm Im}(\omega) \to -\infty$; it thus satisfies the properties required for the derivation of the system of ordinary differential equations (\ref{eqforrst}) for the evolution of the order parameter. On the contrary, in the lower half plane the
function (\ref{finit2}) has a pole in $\omega = -\ii$. For both choices of initial conditions $f(\omega,0)$ it is not difficult to compute
the initial value of the order parameter, from Eq. (\ref{orderfour_1}), when $g(\omega)$ is the Gaussian. In fact, we have
\begin{equation}
\label{r0in1}
R(0)=r(0)= \frac{1}{5\sqrt{\pi}} \int_{-\infty}^{+\infty} \dd \omega \, e^{-\omega^2 - \ii \omega} = \frac{1}{5}e^{-0.25} \approx 0.15576
\ee
for $f(\omega,0)$ given by (\ref{finit1}), and, using a known result,
\be
\label{r0in2}
R(0) = r(0) = \frac{\delta}{\sqrt{\pi}} \int_{-\infty}^{+\infty} \dd \omega \, \frac{e^{-\omega^2}}{\omega^2+1} = 
\delta \sqrt{\pi} e \left[ 1 - {\rm erf}(1)\right] \, ,
\ee
where ${\rm erf}(x)$ is the error function, for $f(\omega,0)$ given by (\ref{finit2}). The value of $\delta$ ($\approx 0.20552$) has been
taken in order to give the same value of $R(0)$ as in (\ref{r0in1}). The order parameter evolution coming from the two simulations of the partial differential equation (\ref{eqneq1}) starting from the two initial conditions is presented below, in Section \ref{secresults2}, in
Fig. \ref{figdynam_1} and Fig. \ref{figdynam_2}, respectively.

We now go to the choice of the initial conditions for the integration of the $N$-body equations of motion (\ref{eq_kura_mod}).
In principle it would be possible to reproduce the initial condition $f(\omega,0)$ also in the integration of the $N$-body
equations of motion, but in practice it is not possible to achieve this with a reasonable precision without simulating a system with a
huge number $N$ of oscillators, in which case the integration would require an unreasonable time. Therefore in the simulations of the
$N$-body system we have chosen, as initial condition, values of the phases $\theta_i$ uniformly distributed in a given range between
$0$ and $2\pi$, with the range chosen to produce the desired initial value $R(0)$. To have the same initial value of
$R(0)$ as in the integration of the partial differential equation (\ref{eqneq1}), the initial phases have been uniformly distributed
between $-\theta_0$ and $+\theta_0$, with $\theta_0 \approx 2.70643$. The resulting dynamics of the order parameter is given both in
Fig. \ref{figdynam_1} and in Fig. \ref{figdynam_2}, for comparison with the corresponding dynamics obtained from the other evolution
equations.

We finally consider the initial conditions for the system of ordinary differential equations (\ref{eqforrst}). In these equations the
complex order parameter $r(t)$ is expressed in terms of the dynamical variables $r^{(m)}(t)$ through Eq. (\ref{residuesb}), with the
coefficients $a_m$ proportional to the residues of the approximating function $g_P(\omega)$ (see Eq. (\ref{defam})) and the frequencies
$\omega_m$ the corresponding poles. We thus need the initial values of the dynamical variables, $r^{(m)}(0)$. Correspondingly to the two
choices of the initial condition $f(\omega,0)$, the values of $r^{(m)}(0)$ have been taken as follows. For $f(\omega,0)$ equal to
(\ref{finit1}) we have taken, according to Eq. (\ref{rpart}), $r^{(m)*}(0)$ equal to $f(\omega_m,0)$, i.e., the value of analytic
continuation of $f(\omega,0)$ in the $m$-pole of the approximating function $g_P(\omega)$. On the other hand, for $f(\omega,0)$ equal
to (\ref{finit2}) Eq. (\ref{residuesb}) is no longer valid, even with a $g(\omega)$ with only poles in the lower half plane, like our
$g_P(\omega)$, since now we have to take into account also the residue at the pole of $f(\omega,0)$. Then, Eq. (\ref{residuesb}) cannot be
directly used to compute the initial value $R(0)$ for the system of equations (\ref{eqforrst}). Therefore, we did use the extension of
Eq. (\ref{finit2}) to complex $\omega$ values, taking $r^{(m)*}(0)$ proportional to $f(\omega_m,0)$, with the constant of
proportionality (i.e., an overall coefficient in front of the sum in (\ref{residuesb})) such to get the same initial value $R(0)$
as before ($\approx 0.15576$). This is a justified procedure, since with the choice of $f(\omega,0)$ equal to Eq. (\ref{finit2}) we want
to study the dynamics of the system (\ref{eqforrst}) in a case in which it would not be allowed, mathematically, to represent with that
system the dynamics generated by Eq. (\ref{eqneq1}) for the order parameter, because the initial condition does not belong to the
manifold ${\cal G}_2$; thus, our choice of the initial conditions for the system (\ref{eqforrst}), with $R(0)$ equal to that resulting
from of Eq. (\ref{eqneq1}), seems the most natural choice.

\subsection{The results of the numerical integrations}
\label{secresults2}
In the numerical integration of the equations of motion (\ref{eq_kura_mod}) we have taken $N=4\times 10^5$ oscillators, and the timestep
has been fixed equal to $\Delta t = 0.001$; the frequencies have been extracted from the Gaussian distribution
(\ref{gaussfreqnorm}).
In the results presented below, we have taken, for the coupling constant, the value $K=\frac{3}{2}K_c$. We have obtained similar results
for other values.

In Fig. \ref{figdynam_1} we show the comparison of the order parameter dynamics obtained from the $N$-body equations of motion
(\ref{eq_kura_mod}) with that obtained from the integration of the partial differential equation (\ref{eqneq1}) with initial condition
$f(\omega,0)=\frac{1}{5}e^{-\ii \omega}$ (see Eq. (\ref{finit1})), and with that obtained from the system of ordinary differential
equations (\ref{eqforrst}).
The initial conditions have been taken as described in Section \ref{secresults1}. The integration of the partial differential equation
(\ref{eqneq1}) has been done discretizing the $\omega$ values in steps of $\Delta \omega = 0.0001$. As already
remarked, the function $f(\omega,0)$ given by Eq. (\ref{finit1}) satisfies the properties required to belong to the
manifold ${\cal G}_2$. We recall that the initial value of $R(0)$ is about $0.15576$.
\begin{figure}[ht]
\begin{center}
\includegraphics[scale=0.5,trim= 0cm 0cm 0cm 0cm,clip]{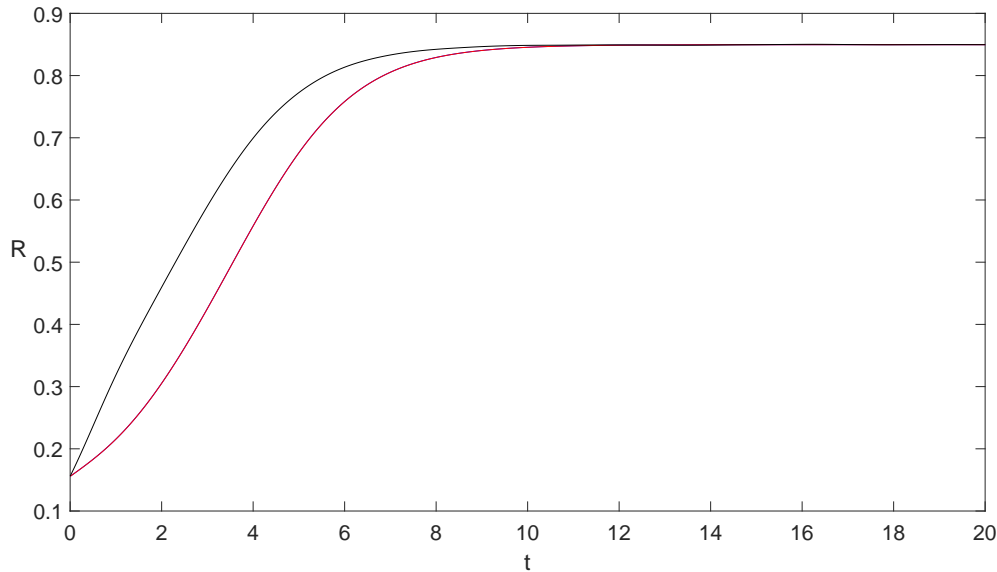}
\caption{Time course of the modulus $R(t)$ of the complex order parameter, for a Gaussian frequency distribution and with
$K=\frac{3}{2}K_c$, from:
solution of the equations of motion (\ref{eq_kura_mod}) (black); solution of Eq. (\ref{eqneq1}) for $f(\omega,t)$ with initial condition
$f(\omega,0) = e^{-\ii \omega}/5$ (blue); solution of the system of equations (\ref{eqforrst}) (red) with the same initial condition
extended to complex frequencies (in this case the Gaussian frequency distribution
is replaced by the approximation $g_{12}(\omega)$). The blue and red curves superimpose, and they are not distinguishable.}
\label{figdynam_1}
\end{center}
\end{figure}

In the plot the curves $R(t)$ for the equation (\ref{eqneq1}) and the system of equations (\ref{eqforrst}) are indistinguishable on the
scale of the plot, while the simulation of the $N$-body system gives a clearly different $R(t)$, that presents a more rapid rise towards
the asymptotic value. Obviously, we have to expect that the initial condition of the system of $N$ oscillators does not satisfy the
OA ansatz and then the $n$-th Fourier component of the initial distribution $\rho(\theta,\omega,0)$ will not be the $n$-th power of the
first one. We come back below to this point, showing a simulation with a Lorentzian frequency distribution. It is interesting to note how
a distribution with poles located at $\omega$ values with moduli of order $1$ (see Fig. \ref{figpoli}) can reproduce the dynamics of a
distribution with an essential singularity at infinite. In the Conclusions and in Appendix \ref{secappend2} we comment on this point.

Next we analyze the case in which the initial condition $f(\omega,0)$ does not belong to the manifold ${\cal G}_2$.
In Fig. \ref{figdynam_2} we compare the order parameter dynamics obtained from the $N$-body equations of
motion (\ref{eq_kura_mod}) (the same dynamics shown in Fig. \ref{figdynam_1}) with that obtained from the integration of the partial
differential equation (\ref{eqneq1}) with initial condition given by Eq. (\ref{finit2}), and with that obtained from the system of
ordinary differential equations (\ref{eqforrst}).
Also for this case we refer, for the initial conditions, to Section \ref{secresults1}.
\begin{figure}[ht]
\begin{center}
\includegraphics[scale=0.5,trim= 0cm 0cm 0cm 0cm,clip]{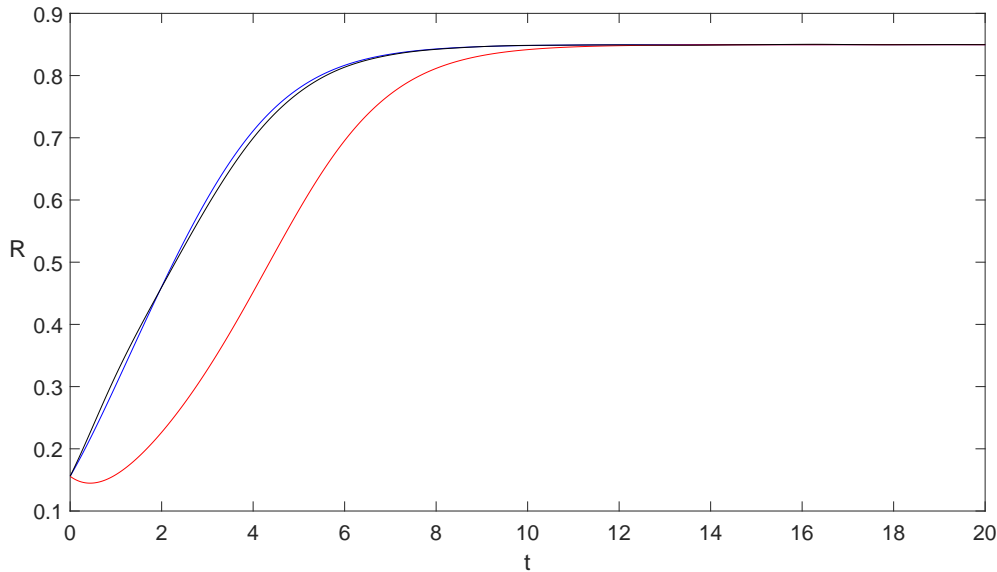}
\caption{Time course of the modulus $R(t)$ of the complex order parameter, for a Gaussian frequency distribution and with $K=\frac{3}{2}K_c$, from:  solution of the equations of motion (\ref{eq_kura_mod}) (black); solution of Eq. (\ref{eqneq1}) for $f(\omega,t)$ with initial
condition $f(\omega,0) = \delta/(\omega^2 +1)$ (blue); solution of the system of equations (\ref{eqforrst}) (red) (in this case the
Gaussian frequency distribution is replaced by the approximation $g_{12}(\omega)$).}
\label{figdynam_2}
\end{center}
\end{figure}

We see that now, as expected, there is a clear difference, in the initial transient, between the $R(t)$ generated by Eq. (\ref{eqneq1}) and
the $R(t)$ generated by the system (\ref{eqforrst}). The former appears close, but clearly different, to the $N$-body
dynamics, for which we use, as remarked, the same dynamics as that in Fig. \ref{figdynam_1}.
Therefore, now the difference, in the initial transient, between the dynamic generated by Eq. (\ref{eqneq1}) and that generated by the
system (\ref{eqforrst}) is due to $f(\omega,0)$ not belonging to the manifold ${\cal G}_2$, while, as before, the difference of both with respect to the $N$-body dynamics is due to the fact that in the latter $\rho(\theta,\omega,0)$ does not satisfy the OA ansatz.

Concerning the last features, we want to show that they occur also in the simple case of a Lorentzian frequency distribution, that clearly
satisfies the requirements for the representation of the dynamics of the order parameter with the system (\ref{eqforrst}), that in this
case consists of only a single equation. Even more, this single equation can be analytically solved \cite{ottant}. It is not necessary here
to give the expression of this analytical solution; our purpose is to compare it with the dynamics of $R(t)$ generated by the $N$-body
equations of motion (\ref{eq_kura_mod}) and with that determined by Eq. (\ref{eqforrst}) for an initial condition $f(\omega,0)$ not
belonging to the manifold ${\cal G}_2$. So, the frequency distribution is $g(\omega)= \frac{b}{\pi}\frac{1}{\omega^2 +b^2}$, where we
have taken $b=0.1$. For $f(\omega,0)$ we have chosen the same function (\ref{finit2}) employed before. Now, the positive factor $\delta$
can be obtained, as a function of $R(0)$, from the expression resulting from Eq. (\ref{orderfour_1}) in this case, which is
$R(0)=r(0)=\delta /(b+1)$. We remind that for the Lorentzian distribution the critical value of the coupling constant is $K_c = 2b$. As
for the previous simulations we have chosen $K=\frac{3}{2}K_c$, and we prepared the system with a finite initial value $R(0) \approx 0.301$.
In Fig. \ref{figdynam_3} we show the comparison of the dynamics generated by the equations of motion (\ref{eq_kura_mod})
with the analytical solution of the single equation (\ref{eqforrst}) when $M=1$ and with the solution of Eq. (\ref{eqneq1}).
\begin{figure}[ht]
\begin{center}
\includegraphics[scale=0.5,trim= 0cm 0cm 0cm 0cm,clip]{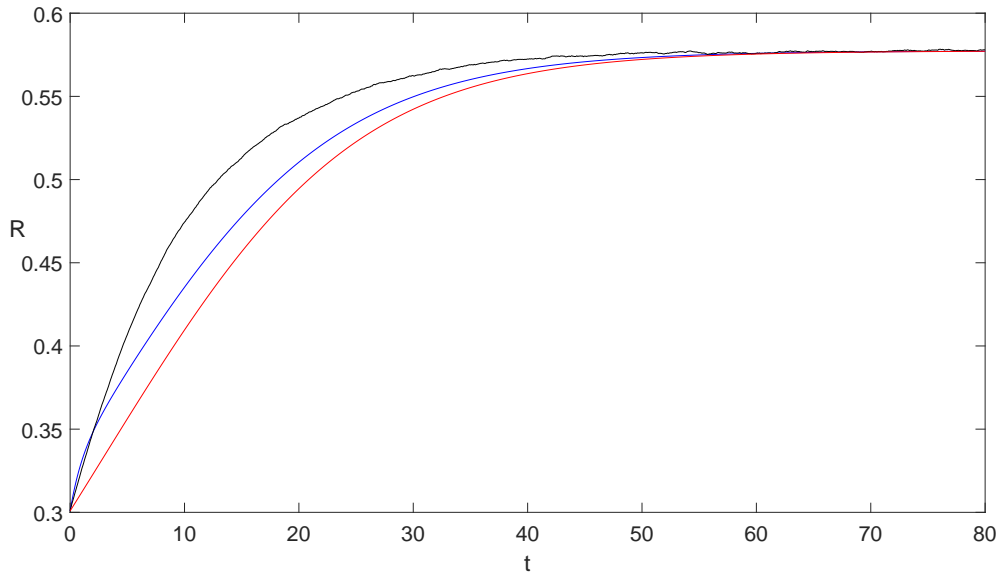}
\caption{Time course of the modulus $R(t)$ of the complex order parameter, for a Lorentzian frequency distribution and with
$K=\frac{3}{2}K_c$, from: solution of the equations of motion (\ref{eq_kura_mod}) (black); solution of Eq. (\ref{eqneq1}) for
$f(\omega,t)$ with initial condition $f(\omega,0) = \delta/(\omega^2 +1)$ (blue); solution of the system of equations (\ref{eqforrst})
(red) (in this case the system reduces to a single equation that can be solved analytically).}
\label{figdynam_3}
\end{center}
\end{figure}
As for the Gaussian frequency distribution there is a clear difference in an initial transient between the various dynamics. The origin
of the differences is the same. Concerning the dynamics generated by the equations of motion (\ref{eq_kura_mod}), it stems from the fact
that $\rho(\theta,\omega,0)$ does not satisfy the OA ansatz, while the difference between the dynamics generated by (\ref{eqneq1}) and
that generated by (\ref{eqforrst}) is a consequence of the initial $f(\omega,0)$, that does not belong to the manifold ${\cal G}_2$.

Overall the results confirm that, with the simple approximation introduced for the treatment of the Gaussian frequency distribution, this
case can be treated, within the scheme of the OA ansatz, analogously to those in which $g(\omega)$ has only few poles in the lower half
plane. The differences with respect to the dynamics determined by the Eq. (\ref{eqneq1}) for $f(\omega,t)$, have therefore the same origin,
as well as the differences with respect to the dynamics described by the equations of motion (\ref{eq_kura_mod}).

\section{Discussion and conclusions}
\label{secdiscuss}
In this work our focus has been the analysis of an approximation that could be suitable for the study of the dynamics of the order
parameter $r(t)$ of the Kuramoto model, in the limit $N \to \infty$, {\it within the OA ansatz}, independently from the
properties of the frequency distribution $g(\omega)$. With the OA ansatz the dynamics, described in general with the continuity
equation (\ref{contequat}) for the distribution function $\rho(\theta,\omega,t)$, can be represented by the partial differential equation
(\ref{eqneq1}), together with Eq. (\ref{orderfour_1}), for the first Fourier component $f(\omega,t)$ of the distribution function
$\rho(\theta,\omega,t)$. Only with further assumptions, especially for the frequency distribution $g(\omega)$, it is possible to go further,
replacing Eq. (\ref{eqneq1}), an equation depending on a continuous variable $\omega$, with the system of (few) ordinary differential
equations (\ref{eqforrst}), if one wants to describe the dynamics of the order parameter. The main purpose of this work has
been the comparison of the long-time dynamics of the order parameter as obtained from the $N$-body equations of motion (\ref{eq_kura_mod})
and from the system of ordinary differential equations (\ref{eqforrst}); in the latter we have employed the poles of the approximating
function $g_P(\omega)$. The numerical equality of the asymptotic value of the order parameter shown by the black and red curves in
Fig. \ref{figdynam_1} and Fig. \ref{figdynam_2}) is the main result of this paper.

Although our work has considered the dynamics in the invariant manifold defined by the OA ansatz, we must now comment about the relation of
this dynamics to the full dynamics as generated by Eq. (\ref{fpstart}) for $\rho(\theta,\omega,t)$ or by Eq. (\ref{systemeq}) for
the Fourier components $f_n(\omega,t)$. As a first remark, we have to stress that, with those equations, we are considering an infinite
system. Therefore {\it in principle} those equations should not be compared with the dynamics of a finite system of oscillators, even with
a very large number of them. We come back to this in a moment, but before we treat the point of the relation of the full dynamics, i.e.,
the system (\ref{systemeq}), with the dynamics in the OA manifold. Although this manifold is invariant, it has been shown \cite{engel2020}
that in general it is not attracting; therefore, if we consider the full dynamics, we should not expect, in general, that the trajectory in
the full dynamical infinite dimensional phase space of the system will converge to the OA manifold; in other words, the trajectory in the
full phase space will in general stay distinct from a trajectory generated by (\ref{eqneq1}), that will remain in the OA manifold. However,
as we have emphasized above, in this paper we are interested in the long-time behavior of the order parameter, not in the analysis of the
trajectory in the full phase space of the system; the long-time behavior of the order parameter is what we need for the determination of
the phase diagram of the system, as explained in the Introduction. This situation can be considered very similar to the study of a
macroscopic observable in a many-body system; we know that the determination of the precise detailed microscopic dynamics, that in general
can be obtained only numerically with simulations, is not necessary for the determination of the macroscopic observable. The analogy
can be further extended by observing that in our oscillator system the order parameter can reach a stationary value while the microscopic
dynamics continues to evolve; this is similar to the relation between the stationary equilibrium value of an observable and the underlying
microscopic dynamics in a many-body Hamiltonian system. Then, for the study of the order parameter in our oscillator system, it suffices
that the dynamics of the order parameter, as generated by the full dynamics, approaches, in some sense, the corresponding dynamics as
generated within the OA manifold. In this respect it is shown in Ref. \onlinecite{engel2020} that for a Lorentzian
$g(\omega)$ and the assumptions on $f(\omega,0)$ required by the OA scheme, this actually occurs in an infinite
system, for a weak convergence to the solution in the OA manifold. Thus, in spite of the fact that a distribution with
$f_n=f_1^n$ is a very special one, Ref. \onlinecite{engel2020} shows that at least for the Lorentzian frequency
distribution this assumption does capture the long time behavior of the order parameter. The extension of the validity
of this result to other functional forms of $g(\omega)$ seems to be very reasonable, but from the mathematical point of view it remains at
the level of an assumption. Our numerical results, however, provide support to it. A concrete system,
in particular the system simulated by numerically solving the equations of motion like (\ref{eq_kura_mod}), has a very large but finite
number $N$ of components. Considering the full microscopic dynamics, it is possible that some features occurring for $N\to \infty$ are no
more valid for finite $N$, even if large. But, again, we are considering here the long-time behavior of a macroscopic quantity, and on
physical grounds one should expect that this quantity in the finite macroscopic system behaves like in the infinite system; in particular,
the asymptotic value of the order parameter could have a relative correction of the order $1/N$ elevated to some positive power with respect
to the one of the infinite system, numerically irrelevant for the $N$ values available in simulations, but sufficient to prevent a
mathematical convergence to the order parameter value for $N \to \infty$. This expectation obviously is not a mathematical proof, and one
can rely on the numerical simulations showing that this is actually what occurs; in our case Fig. \ref{figdynam_1} and Fig. \ref{figdynam_2}.

The numerical convergence of the order parameter as obtained by the system of equations (\ref{eqforrst}) to the one derived from the
$N$-body equations of motion seems to occur also in spite of another issue that, from the mathematical point of view, has to be considered.
As shown in Ref. \onlinecite{engel2020}, the manifold ${\cal G}_1$ is only one of the invariants of the dynamics. There are infinite other
invariant manifolds, defined by finite weighted and normalized sums of different distributions $\rho(\theta,\omega,t)$, but with each one
satisfying the position (\ref{ansoadef}). Furthermore, all these invariant manifolds are only neutrally stable and continuously connected
(full details and formal developments are found in Ref. \onlinecite{engel2020}). It is then to be expected that the numerical simulation of
the dynamics will drift among the different manifolds. This is a common situation in simulations: because of the finite precision due to the
finiteness of the integration step, the orbit determined by the integration algorithm will never be mathematically a true orbit of the
system. Then, in presence of continuously connected neutrally stable invariant manifolds, even if the true orbit remains in one of them,
the simulated orbit, because of the finite precision, will in general drift among the invariant manifolds. In our case, in principle this
fact raises a question about the representativity of the asymptotic values of the order parameter. On the other hand, the reproducibility of
this asymptotic value when one performs simulations starting from different initial conditions, could suggest that, after an initial
transient, the order parameter, {\it in the weak convergence sense mentioned above}, tends to the same asymptotic value regardless of the
invariant manifold to which the true orbit belongs, giving support to the representativity of the asymptotic value. Also in this case, as before,
from the mathematical point of view this remains an assumption, although supported by the numerical results.

As a working example for our study, we have taken a Gaussian frequency distribution. This very natural assumption about the form of
$g(\omega)$ does not satisfy the necessary conditions for the use of the system (\ref{eqforrst}). However, on the basis of the approximating
function that we have presented, we have shown that, from the practical point of view, Eqs. (\ref{eqforrst}) can actually be employed.
It is interesting to look for an explanation of why an essential singularity, like that of the Gaussian frequency distribution, while
preventing mathematically the use of the system (\ref{eqforrst}), nevertheless does not have any numerically relevant consequence on the
asymptotic dynamics of the order parameter. In Appendix \ref{secappend2} we show that this is related to the fact that, for
the frequency distribution proportional to $e^{-\omega^2}$, the essential singularity is associated to a term of the form
$e^{-t^2/4}$ in the dynamics of $r(t)$; such a term gives a numerically negligible contribution, especially in the case in which
$|r(t)|$ tends asymptotically to a finite value. However, being always present, independently from the properties of the analytical
continuation to complex values of the initial condition $f(\omega,0)$, it does not allow to have a mathematically rigorous
description with a system of equations like (\ref{eqforrst}). For other frequency distributions with an essential singularity
at infinite, the coefficient of the exponent will be different, i.e., we will have a term of the form $e^{-\gamma t^2}$, with
$\gamma$ positive, but this does not change the conclusion.

We have compared the dynamics of the order parameter derived from the numerical solution of (\ref{eqneq1}) and of (\ref{eqforrst}) with
the dynamics of the order parameter as obtained in the simulation of a $N$-body system, i.e., of the equations of motion (\ref{eq_kura_mod}).
We have seen that at early times the dynamics do not coincide, but that very soon they do so, in particular tending to the same asymptotic
value for $|r(t)|$. We then see that the value chosen for $N$, i.e., $4\times 10^5$, is sufficient to have a dynamics {\it of the order
parameter} that is practically coincident with that of an infinite system. To use the same argument employed above, this is analogous to
what found in the $N$-body simulations of macroscopic systems, obviously with a finite $N$, that agree with the results obtained, for the
macroscopic observables, with theoretical analyses made in the thermodynamic limit $N \to \infty$.

Concerning the differences in the initial transient of the dynamics of the order parameter, we have shown that they are unavoidable, also in
the case in which the frequency distribution allows the reduction of the dynamics, for suitable initial conditions $f(\omega,0)$, to the
low dimensional manifold ${\cal M}$. In fact, they appear in the simplest such case, that of a Lorentzian, with only a single
pole in the lower half plane (see Fig. \ref{figdynam_3}). Therefore the differences are present in general, and our approximation
scheme, while allowing, {\it within the OA ansatz}, the reduction of the order parameter dynamics to the low dimensional manifold ${\cal M}$,
cannot remove them. Also, the differences in the transient behavior will depend on the initial conditions, but if one is
interested in the asymptotic behavior of the order parameter this will not be a problem.

The practical value of the results presented in this work could be the following. Here we have analyzed the results with a Gaussian
frequency distribution. As emphasized above, for symmetric unimodal frequency distribution the phase diagram is quite simple. But for
other distributions the phase diagram can be very rich. Especially for non-unimodal distributions we could expect the presence
not only of stationary states for the modulus $R(t)$ of the order parameter, but also of oscillating asymptotic states \cite{bimodal}.
The approximation of a frequency distribution not satisfying the assumptions for the use of the system (\ref{eqforrst}) with one that
does satisfy them, could then be exploited for the determination of the phase diagram in a fast way. In fact, the numerical solution
of the system (\ref{eqforrst}) is extremely faster than that of Eq. (\ref{eqneq1}) or of the simulation of the equations of motion
(\ref{eq_kura_mod}) of the $N$-body system. We do recognize that the borders between different regions in the phase diagram will be
determined with a small error. For example, the transition point in the onedimensional phase diagram of the Gaussian frequency distribution
is affected by a relative error of about $C_P -1$, where $C_P$ is the normalizing factor of the approximating function
(\ref{apprgauss}). However, we have seen that $C_P$ is extremely close to $1$ (we remind that $C_{12}\approx 1.00003$).
The application of the approximation scheme proposed here to other distributions would require that, similarly
to what found for the Gaussian, the normalizing factor $C_P$ would be very close to $1$. We remind that this closeness is here related
to the fact that the fraction in the approximating function (\ref{apprgauss}) is numerically so close to $e^{-\omega^2}$ that its
integral is extremely close to $\sqrt{\pi}$. Therefore for the application of the scheme one should find an analogous relation
between the approximating function and the original function not satisfying the assumption for the use of the system
(\ref{eqforrst}). The number of necessary poles in the approximating function might sometimes be higher than the one found here for the
Gaussian, but this would not be important, given the extreme velocity of the numerical solution of the system (\ref{eqforrst}).
We therefore expect that also in the application to other cases the asymptotic states of the system, and their properties,
will be found correctly, with just this very small error in the location of the borders between different regions.

An intrinsic limitation in the OA procedure, more precisely in the use of a system of few ordinary differential equations like
(\ref{eqforrst}), comes from the requirement that the frequency distribution function $g(\omega)$ be analytically prolongable in the complex
plane. This excludes, for example, continuous distributions that however have a compact support in the real $\omega$ line. But also in this
case, if one is able to find an approximating function which is continuous and analytic on the whole real line, and with few poles in the
lower half plane, the method proposed here could be applied.

This work has focused on the study of the phase diagram of the system of oscillators, adopting the point of view of a statistical mechanics study. In this perspective, the object of interest is a macroscopic quantity, here the order parameter, and the representation of the
dynamics with the system (\ref{eqforrst}), i.e., the use of the OA ansatz, can be interpreted as an approximation for the dynamics of the
order parameter. Of course the usefulness of the approximation represented by the OA ansatz resides in the fact, as shown by the
simulations, that the long-time behavior of the order parameter is caught correctly. The approximation scheme cannot be adopted if the
frequency distribution $g(\omega)$ has essential singularities in the complex plane, and this work has dealt with a suitable approximation
of $g(\omega)$ that allows to use it. If other macroscopic quantities happen to share with the order parameter $r(t)$ the property that
their long-time behavior can be well represented within the OA ansatz, then the method presented in this work could possibly be adopted
also in their study.

\begin{acknowledgments}
The author acknowledges financial support from INFN (Istituto Nazionale di Fisica Nucleare) through the projects DYNSYSMATH and ENESMA.
\end{acknowledgments}

\section*{Author declaration}
The author has no conflicts to disclose.

\section*{Data availability}
The data that support the findings of this study are available within the article.

\appendix
\section{Some details on the solution of Eq. (\ref{eqneq1lin})}
\label{secappend1}
Here we provide some details about the solution of the linearized equation (\ref{eqneq1lin}), in particular concerning the computation
of the integrals in (\ref{neededint}), when the frequency distribution is the Lorentzian
\be
\label{gloren_app}
g(\omega) = \frac{b}{\pi}\frac{1}{\omega^2 + b^2} \, .
\ee
We begin by computing the second integral in Eq. (\ref{neededint}), i.e.,
\be
\label{neededint_second}
\int \dd \omega \, \frac{g(\omega)}{\omega -\nu} = \frac{b}{\pi} \int \dd \omega \, \frac{1}{(\omega - \nu )(\omega^2 + b^2)} \, .
\ee
Remembering that performing this integral we have to assume ${\rm Im}(\nu)>0$, we obtain, applying the residue theorem
and noting that the integrand has a pole, in the lower half plane, at $\omega = -\ii b$,
\bea
\label{expr_app_1}
&&\frac{b}{\pi} \int \dd \omega \, \frac{1}{(\omega - \nu )(\omega^2 + b^2)}
= -2\pi \ii {\rm Res} \left. \left[\frac{b}{\pi}\frac{1}{(\omega - \nu )(\omega^2 + b^2)}\right] \right|_{\omega=-\ii b}
\nonumber \\
&=& -2 \ii b \left. \left[\frac{1}{(\omega - \nu )(\omega -\ii b)}\right] \right|_{\omega=-\ii b}
= - \frac{1}{\nu +\ii b} \, ,
\eea
which is Eq. (\ref{value_int2}). The expression (\ref{specdiel}) for the function $\epsilon(\nu)$ is obtained by substituting
in Eq. (\ref{defindiel}). Going now to the second integral in (\ref{neededint}), i.e.,
\be
\label{neededint_first}
\int \dd \omega \, \frac{f(\omega,0)g(\omega)}{\omega-\nu}
= \frac{b}{\pi} \int \dd \omega \, \frac{f(\omega,0)}{(\omega - \nu )(\omega^2 + b^2)} \, ,
\ee
we first consider the case in which $f(\omega,0)$ has no singularities in the lower half plane. Then, from the previous integral
one obviously deduces that
\be
\label{expr_app_2}
\frac{b}{\pi} \int \dd \omega \, \frac{f(\omega,0)}{(\omega - \nu )(\omega^2 + b^2)}
= - \frac{f(-\ii b,0)}{\nu +\ii b} \, .
\ee
On the other hand, when $f(\omega,0)$ is the Lorentzian (\ref{fomega0def}), we have
\bea
\label{expr_app_3}
&&\frac{bc}{\pi^2} \int \dd \omega \, \frac{1}{(\omega - \nu )(\omega^2 + b^2)(\omega^2 + c^2)}
= -2\pi \ii \frac{bc}{\pi^2}\left\{ {\rm Res} \left.
\left[\frac{1}{(\omega - \nu )(\omega^2 + b^2)(\omega^2 + c^2)}\right] \right|_{\omega=-\ii b}
\right. \nonumber \\ &&\,\,\,\,\,\,\,\,\,\,\,\,\,\,\,\,\,\,\,\,\,\,\, \left. +
{\rm Res} \left.
\left[\frac{1}{(\omega - \nu )(\omega^2 + b^2)(\omega^2 + c^2)}\right] \right|_{\omega=-\ii c} \right\}
\\
&=& -\frac{2\ii bc}{\pi} \left\{ \left.
\left[\frac{1}{(\omega - \nu )(\omega^2 - \ii b)(\omega^2 + c^2)}\right] \right|_{\omega=-\ii b} +
\left. \left[\frac{1}{(\omega - \nu )(\omega^2 + b^2)(\omega^2 - \ii c)}\right] \right|_{\omega=-\ii c} \right\}
\, . \nonumber
\eea
From the expression in the last member one easily gets the expression in the right hand side of Eq. (\ref{value_int1_b}).
Now, with the help of the expression for $\epsilon(\nu)$, Eq. (\ref{specdiel}), those appearing in Eqs. (\ref{rnu_1}) and
(\ref{rnu_2}) are obtained using, respectively, (\ref{expr_app_2}) and (\ref{expr_app_3}). It is not necessary
to describe the simple calculation, along the same line, of $r^*(0)$.

The expressions (\ref{roft_1}) and (\ref{roft_2}) of $r^*(t)$ are obtained from the expressions (\ref{rnu_1})
and (\ref{rnu_2}), respectively, of $\widetilde{r}^*(\nu)$, by noting that for $t>0$, applying the residue theorem
we have
\be
\label{invlapgen}
\frac{1}{2\pi} \int_{{\cal C}} \dd \nu \, e^{-\ii \nu t} \frac{h(\nu)}{\nu - \nu_0}
= -\ii h(\nu_0) e^{-\ii \nu_0 t} \,
\ee
with the path ${\cal C}$ passing above $\nu_0$ and with the function $h(\nu)$ having no singularities in the region
below ${\cal C}$.

\section{Computation of $\widetilde{r}^*(\nu)$ as given by Eq. (\ref{laptransorder_1}) for a Gaussian
frequency distribution}
\label{secappend2}
Here we compute the expression in the right hand side of Eq. (\ref{laptransorder_1}) when $g(\omega)$ is the Gaussian
(\ref{gaussfreqnorm}) and when the initial condition $f(\omega,0)$ is equal to $\beta e^{-\ii \omega}$,
with $\beta$ a positive number smaller than $1$.
Thus, we are in the situation in which the system of equations (\ref{eqforrst_lin}) cannot represent the dynamics of $r^*(t)$
as determined by Eq. (\ref{eqneq1lin}), but we choose an $f(\omega,0)$ that can be continued in the lower half plane
without singularities and tending to zero for ${\rm Im}(\omega) \to -\infty$.

We proceed with the computation of the integrals in (\ref{neededint}), needed for the evaluation of the right hand side
of Eq. (\ref{laptransorder_1}). Let us begin with
\be
\label{intgauss1}
\frac{1}{\sqrt{\pi}} \int \dd \omega \, \frac{e^{-\omega^2}}{\omega - \nu} \, ,
\ee
for ${\rm Im}(\nu) >0$, needed for the computation of $\epsilon(\nu)$. We introduce the function
\be
\label{intgauss1_a}
A(\alpha,\nu) = \frac{1}{\sqrt{\pi}} \int \dd \omega \, \frac{e^{-\alpha \omega^2}}{\omega - \nu} \, ,
\ee
again for ${\rm Im}(\nu) >0$, and with $\alpha$ a real positive number. Our integral is clearly equal to $A(1,\nu)$.
Multiplying for $1= (\omega + \nu)/(\omega + \nu)$ we get
\be
\label{intgauss1_b}
A(\alpha,\nu) = \frac{\nu}{\sqrt{\pi}} \int \dd \omega \, \frac{e^{-\alpha \omega^2}}{\omega^2 - \nu^2} \, .
\ee
We note that the denominator, as in the original expression, never vanishes for real $\omega$. By deriving this
expression with respect to $\alpha$ we obtain in a straightforward way that
\be
\label{intgauss1_c}
\frac{\partial}{\partial \alpha}A(\alpha,\nu) = -\frac{\nu}{\sqrt{\alpha}} - \nu^2 A(\alpha,\nu) \, .
\ee
Noting that $a(0,\nu) = \ii \sqrt{\pi}$, the solution of this linear equation is readily obtained as
\be
\label{intgauss1_d}
A(\alpha,\nu) = \sqrt{\pi} e^{-\nu^2} \left[ \ii - \frac{2}{\sqrt{\pi}} \int_0^{\nu\sqrt{\alpha}} \dd z \,
e^{z^2} \right] \, ,
\ee
where the integral in the right hand side is the error function evaluated at the complex value $\nu\sqrt{\alpha}$.
In fact, with the same procedure one can compute the more common integral analogous to (\ref{intgauss1_b}), but in which
the denominator is $(\omega^2 + \lambda^2)$, with $\lambda$ a given real number.
Substituting $\alpha = 1$ we have
\be
\label{intgauss1_e}
\frac{1}{\sqrt{\pi}} \int \dd \omega \, \frac{e^{-\omega^2}}{\omega - \nu} =
A(1,\nu) = \sqrt{\pi} e^{-\nu^2} \left[ \ii - \frac{2}{\sqrt{\pi}} \int_0^{\nu} \dd z \,
e^{z^2} \right] \, ,
\ee
The expression on the right hand side is the analytic continuation in the whole complex $\nu$ plane of the integral
in the left hand side, valid for ${\rm Im}(\nu) >0$. It can be seen that, like the integral in the left hand side,
the right hand side tends to zero as $1/\nu$ for ${\rm Im}(\nu) \to +\infty$. However, for ${\rm Im}(\nu) \to -\infty$
the right hand side, for the presence of an essential singularity, diverges as $e^{[{\rm Im}(\nu)]^2}$.

We now evaluate the other integral in (\ref{neededint}). With the choice made for $f(\omega,0)$ we have to
compute the integral
\be
\label{intgauss2}
\frac{\beta}{\sqrt{\pi}} \int \dd \omega \, \frac{e^{-\omega^2 -\ii \omega}}{\omega - \nu} \, ,
\ee
It is not difficult to see that this integral is equal to $\beta e^{-\frac{1}{4}}A\left(1,\nu + \frac{\ii}{2}\right)$; therefore
we have
\be
\label{intgauss2_a}
\frac{\beta}{\sqrt{\pi}} \int \dd \omega \, \frac{e^{-\omega^2 -\ii \omega}}{\omega - \nu} =
\beta\sqrt{\pi} e^{-\nu^2 -\ii \nu} \left[ \ii - \frac{2}{\sqrt{\pi}} \int_0^{\nu+\frac{\ii}{2}} \dd z \,
e^{z^2} \right] \, .
\ee
Like before, the expression in the right hand side diverges as $e^{[{\rm Im}(\nu)]^2}$ for ${\rm Im}(\nu) \to -\infty$.

We can now use Eqs. (\ref{intgauss1_e}) and (\ref{intgauss2_a}) for the computation of $\widetilde{r}^*(\nu)$ from
Eq. (\ref{laptransorder_1}). The integral (\ref{intgauss1_e}) is used to have $\epsilon(\nu)$ from (\ref{defindiel}).
We do not write explicitly the resulting expression, but we note the following: plugging it in the right hand
side of Eq. (\ref{invtogetr}) to get $r^*(t)$, the result will be determined by the zeros of $\epsilon(\nu)$ and by the
essential singularity of (\ref{intgauss2_a}) for ${\rm Im}(\nu) \to -\infty$. From Eq. (\ref{defindiel}) we see that
in the case we are considering we have $\epsilon(\nu) = 1 + \ii \frac{K}{2}A(1,\nu)$. This expression will vanish
for a value $\nu = \nu_K$ depending on $K$, giving rise to term in $r^*(t)$ going like $e^{-\ii \nu_K t}$. Another
contribution will come from the essential singularity at ${\rm Im}(\nu) \to -\infty$. We want to show that this
essential singularity is related to a time dependence of $r^*(t)$ of the form $e^{-\frac{t^2}{4}}$. In fact,
let us evaluate the Laplace transform of the latter function, i.e.,
\be
\label{laptsquare}
\int_0^{+\infty} \dd t \, e^{\ii \nu t} e^{-\frac{t^2}{4}} \, .
\ee
that for convenience we denote as $B(\nu)$. This integral is not as trivial as the analogous one in which the lower limit
of integration is $-\infty$. We can however compute it in the following way. Making the derivative with respect to $\nu$
we obtain
\be
\label{derlaptsquare}
\frac{\dd B}{\dd \nu} = 2\ii -2\nu B \, .
\ee
This equation is easily solved, noting that $B(0)=\sqrt{\pi}$, as
\be
\label{laptsquare_a}
B(\nu) = \sqrt{\pi} e^{-\nu^2} \left[ 1 + \frac{2\ii}{\sqrt{\pi}} \int_0^{\nu} \dd z \, e^{z^2} \right] \, .
\ee
Like the expression in the right hand side of Eq. (\ref{intgauss2_a}), this one has an essential singularity at infinity,
vanishing for ${\rm Im}(\nu) \to +\infty$ and diverging as $e^{[{\rm Im}(\nu)]^2}$ for ${\rm Im}(\nu) \to -\infty$. This
proves that the essential singularity is caused by a term in $r^*(t)$ of the form $e^{-\frac{t^2}{4}}$. In turn, this
shows that the reason for which the system of Eqs. (\ref{eqforrst_lin}) cannot represent the dynamics of $r^*(t)$
as determined by Eq. (\ref{eqneq1lin}), is in the fact that a time dependence like this is always present, independently from
the initial condition $f(\omega,0)$. Such time dependence cannot be reproduced by the system (\ref{eqforrst_lin}),
that, having a finite number of equations, can give rise only to exponents linear in the time $t$. Although all these
considerations have been based on the linearized equations, they are valid also for the full nonlinear equation
(\ref{eqneq1}), since for very small $f(\omega,t)$ the nonlinear system reduces to the linear one.



\begin{thebibliography}{50}
\bibitem{footnote1} If the parameters characterizing the distribution $g(\omega)$ are $s$, we expect that the dynamics
depends on the ratio between these parameters and the coupling constant $K$, so that the phase diagram has dimension $s$.
\bibitem{footnote2} More rigorously, the continuity equation should be written for the $N$-particle distribution function,
but the factorization approximation usually employed for systems with mean-field interactions, when $N \to \infty$,
allows to get an equation for the one-particle distribution function. For the passage from the $N$-particle
distribution function to the one-particle distribution function, even in the more general context in which
the system is subject to noise and the description has to be made with a Fokker-Planck equation, see, e.g.
Ref. \onlinecite{briefbook}.
\bibitem{footnote3} It is not necessary to make also for $g(\omega)$ the assumption that
$|g(\omega)| \to 0$ for ${\rm Im}(\omega) \to -\infty$, since this is a consequence of the fact that it has no
singularities in the lower half plane other than some poles at finite values; together with the physical
requirement that $g(\omega)$ vanishes for $\omega$ real tending to $\pm \infty$, this implies that it has to
vanish also for ${\rm Im}(\omega) \to -\infty$.
\bibitem{footnote4} A linearized Vlasov equation would have, as a last term, something proportional
to $g(\omega) \int \dd \omega' \, f(\omega',t)$, i.e., with the known function $g(\omega)$ outside the integral.
\bibitem{footnote5} We could not avoid to use the universally adopted symbol $\omega$ for the proper frequencies of
the oscillators; thus, we had to use another symbol, $\nu$, for the variable conjugated to the time $t$. This should
not generate confusion.
\bibitem{footnote6} For the case $M=1$ the solution of the single equation, to which the system (\ref{eqforrst}) reduces
to, is known analytically also for the full nonlinear case\cite{ottant}, but here we consider only the linearized
equation, since we want to compare with the result of the linear equation (\ref{eqneq1lin}).
\end{thebibliography}
\end{document}